\documentclass[a4paper,fleqn]{cas-dc}
\usepackage{float}
\usepackage{stfloats}
\usepackage{times}
\usepackage{graphicx}
\usepackage{amsmath} 
\usepackage{amssymb}
\usepackage{mathtools}
\usepackage{tabularx}
\usepackage[english]{babel}
\usepackage[natbibapa]{apacite}
\usepackage{csquotes}
\usepackage[figurename=Fig.]{caption}
\usepackage[skip=0pt]{caption}

\usepackage{hyperref}

\hypersetup{colorlinks,allcolors=cyan}

\newcommand{\stkout}[1]{\ifmmode\text{\sout{\ensuremath{#1}}}\else\sout{#1}\fi}
\def\tsc#1{\csdef{#1}{\textsc{\lowercase{#1}}\xspace}}
\tsc{WGM}
\tsc{QE}
\tsc{EP}
\tsc{PMS}
\tsc{BEC}
\tsc{DE}

\begin{document}
\let\WriteBookmarks\relax
\def\floatpagepagefraction{1}
\def\textpagefraction{.001}
\shorttitle{\emph{Leveraging Social Media and Google Trends to Identify Waves of Avian Influenza Outbreaks in USA and Canada}}
\shortauthors{M Soltani et~al.}

\title [mode = title]{Leveraging Social Media and Google Trends to Identify Waves of Avian Influenza Outbreaks in USA and Canada} 

\author[1]{Marzieh Soltani}
\fntext[fn1]{soltanik@uoguelph.ca}
\fnmark[1]

\author[1]{Rozita Dara}[orcid = 0000-0002-3728-0275]
\fnmark[2]
\address[1]{School of Computer Science, University of Guelph, Guelph, Ontario, Canada}

\cormark[1]
\ead{drozita@uoguelph.ca}
\cortext[cor1]{Corresponding author}

\fntext[fn2]{drozita@uoguelph.ca}

\author[2]{Zvonimir Poljak}
\address[2]{Department of Population Medicine, Ontario Veterinary College, University of Guelph, Guelph, Ontario, Canada}
\fntext[fn3]{zpoljak@uoguelph.ca}
\fnmark[3]

\author[3]{Caroline Dubé}
\address[3]{Canadian Food Inspection Agency, Animal Health Risk Assessment, Ottawa, Ontario, Canada}
\fntext[fn4]{caroline.dube@inspection.gc.ca}
\fnmark[4]

\author[1]{Neil Bruce}
\fntext[fn5]{brucen@uoguelph.ca}
\fnmark[5]

\author[4]{Shayan Sharif}
\address[4]{Department of Pathobiology, University of Guelph, Guelph, Ontario, Canada}
\fntext[fn6]{shayan@uoguelph.ca}
\fnmark[6]

\begin{abstract}
Avian Influenza Virus (AIV) poses significant threats to the poultry industry, humans, domestic animals, and wildlife health worldwide. Monitoring this infectious disease is important for rapid and effective response to potential outbreaks. Conventional avian influenza surveillance systems have exhibited limitations in providing timely alerts for potential outbreaks. This study aimed to examine the idea of using online activity on social media, and Google searches to improve the identification of AIV in the early stage of an outbreak in a region. To this end, to evaluate the feasibility of this approach, we collected historical data on online user activities from X (formerly known as Twitter) and Google Trends and assessed the statistical correlation of activities in a region with the AIV outbreak officially reported case numbers. In order to mitigate the effect of the noisy content on the outbreak identification process, large language models were utilized to filter out the relevant online activity on X that could be indicative of an outbreak. Additionally, we conducted trend analysis on the selected internet-based data sources in terms of their timeliness and statistical significance in identifying AIV outbreaks.  {Moreover, we performed an ablation study using autoregsressive forecasting models to identify the contribution of X and Google Trends in predicting AIV outbreaks.} The experimental findings illustrate that online activity on social media and search engine trends can detect avian influenza outbreaks, providing alerts earlier compared to official reports. This study suggests that real-time analysis of social media outlets and Google search trends can be used in avian influenza outbreak early warning systems, supporting epidemiologists and animal health professionals in informed decision-making.
\end{abstract}

\begin{keywords}
X \sep  Social Media \sep Google Trends \sep Avian Influenza \sep Early Identification \sep Large Language Models \sep Public Health Surveillance
\end{keywords}

\maketitle

\section{Introduction}
\label{intro}

Avian Influenza Virus (AIV) poses a significant threat, causing substantial agricultural economic losses due to extensive poultry mortality following outbreaks. For instance, the 2014–2015 Highly Pathogenic Avian Influenza (HPAI) epidemic in the USA stands as one of the largest recorded, resulting in approximately \$3.3 billion in direct production losses, along with an extra \$610 million in federal government response costs \citep{seeger2021poultry, johnson2016local}. Since 2021, there has been a widespread presence of HPAI H5 subtype viruses globally, affecting both avian and mammalian species, resulting in substantial financial hardships \citep{xie2023episodic}. AIV also poses a serious public health concern due to its ability to transmit from avian hosts to mammals, including humans. Over the past two decades, various AIV subtypes have caused infections in humans with different clinical symptoms ranging from mild to severe \citep{yang2022human, abubakar2023avian, blagodatski2021avian}. Although human-to-human transmission remains limited, the potential for AIV subtypes such as H5N1 and H7N9 to spark influenza pandemics is a significant concern \citep{philippon2020avian}. The escalated infection rates among mammals in recent avian influenza epidemics also draw attention to the potential of HPAI to pose an increased risk to humans \citep{worldongoing, leguia2023highly, zhao2019semiaquatic}. 

Given the highly contagious nature of AIV, it is crucial to monitor the emergence of such viruses to detect and respond to potential epidemics. To this end, epidemic intelligence has been utilized to help authorities and decision-makers react promptly to such disease emergencies and thereby reduce or eliminate the consequences \citep{christaki2015new, arinik2023evaluation}. Epidemic intelligence is a process that focuses on identifying, investigating, and monitoring potential health threats \citep{paquet2006epidemic}. It involves gathering information from both formal and verified sources, like official reports, as well as informal and unverified sources such as the web, e.g., social networks, search engine trends, and blogs. With millions of global users, these platforms enable real-time sharing of health-related information, providing researchers with valuable data for public health surveillance. Research has shown that combining these sources improves the sensitivity of surveillance systems, improving the overall effectiveness \citep{kaiser2006epidemic, bahk2015comparing, barboza2013evaluation, dion2015big}.

Previous research has also shown the potential of social media platforms and search engines in detecting outbreaks, including avian influenza \citep{paul2014twitter, culotta2010towards, fast2018predicting, zhang2022intelligent, liu2021forecasting}. The web serves as a critical source of health information for many who search for disease-related information online. Consequently, search engines can serve as a new means for disease surveillance. Google introduced the first online disease tracking tool in which Google search query data was used to detect influenza outbreaks for humans. The tool utilizes statistical methods such as linear regression to forecast weekly influenza trends in the US \citep{ginsberg2009detecting}. Since then, the utilization of Google search engine data in population health studies has expanded. Specifically, various studies have focused on early identification of outbreaks, which involves identifying an increase in occurrences prior to the official reports, across a spectrum of infectious diseases such as COVID-19, Ebola, Lyme disease, and Zika \citep{yousefinaghani2021prediction, graham2018prepared, kapitany2019can, morsy2018prediction}.

Studies have also been exploring the possibility of using health-related information found on social media platforms like X for public health surveillance. \cite{culotta2010towards} employed different methods to identify H1N1 influenza-related content on X, comparing them to official Centers for Disease Control and Prevention (CDC) statistics using over 500k posts. The study found that employing a text classifier, particularly through multiple linear regression on posts could achieve a correlation similar to CDC statistics. This implies the possibility of leveraging such models to identify relevant messages. \cite{dimartino2017} utilized the Early Aberration Reporting System algorithms for outbreak detection and proposed a method to verify Twitter alerts by identifying relevant events in unstructured ProMED-mail documents based on identified medical conditions and geographic references. \cite{santillana2015combining} developed a machine learning method that combines data from Google searches, X, hospital records, and participatory surveillance for real-time and forecast estimates of influenza activity in the US. Their method outperformed individual data sources, accurately predicting flu activity up to four weeks ahead of official reports, showing the benefits of using diverse data streams like social media for better flu predictions. \cite{ahmed2018using} analyzed X data from peak periods of the 2009 H1N1 pandemic and 2014 Ebola outbreaks to understand public perceptions for health authority guidance.

Despite the evident benefits outlined in previous studies emphasizing the potential of monitoring online activity in outbreak identification \citep{moorhead2013new, bernardo2013scoping, covidprediction, alkouz2022deepluenza, deiner2024use}, there is limited avian influenza-related research work at the time of our study. \cite{chen2019avian} investigated internet search and social media data for tracking H7N9 avian influenza in China. They found a link between search queries and H7N9 outbreaks, suggesting these could predict outbreaks before they happen. \cite{yousefinaghani2019assessment} developed a system using X data to track avian influenza outbreaks by filtering relevant posts and analyzing over 200K tweets from 2017 to 2018. They examined X's potential to reflect official outbreak reports and found that 75\% of real outbreaks were detectable. This study illustrated the feasibility of using X posts generated as a supplementary tool for monitoring avian influenza. In order to label relevant posts, the authors used a manually labeled sample to train a Naive Bayes semi-supervised learning algorithm to assign labels to the unlabeled data and then built an Internet-based tracking system.

The use of social media and online sources for disease prediction has not been fully explored for early identification of animal-related infectious disease outbreaks such as avian influenza. Previous studies lacked detailed exploration due to their shorter timeframes using limited historical data. Additionally, due to the longer duration and larger extent of the ongoing recent HPAI outbreaks, further research in this area is required. Moreover, existing studies on avian influenza mostly lacked a form of geographical filtering that would allow region-specific analysis for outbreak identification. Another underexplored aspect is comparing X and Google Trends data in detecting avian influenza outbreak waves in a timely manner. Moreover, irrelevant social media content creates noise that can affect outbreak identification and increase computational costs. Filtering out these irrelevant content can improve the accuracy of outbreak identification. Most studies have relied on traditional machine learning approaches such as Latent Dirichlet Allocation or Naïve Bayes \citep{robertson2016avian, yousefinaghani2019assessment} to find relevant online activities that could be indicative of an outbreak. They have not utilized state-of-the-art text classification solutions, e.g., Large Language Models (LLM) \citep{bommasani2021opportunities}, which could be far more effective in rooting out irrelevant posts.

The present study aims to assess the feasibility of an early identification mechanism for avian influenza outbreaks by monitoring online activity on social media and search engines. Specifically, we leveraged geolocated historical data gathered from X and Google Trends.  {Recognizing the limitations of keyword-matching techniques, which often fail to capture the nuances of language, a fine-tuned LLM was employed during the preprocessing of X posts to filter out irrelevant content and identify posts that reflected early signals of outbreaks. This step ensured a more accurate dataset by removing irrelevant or misleading content associated with the keywords.}

To examine if online activity could be used as an early indicator of an outbreak, the cross-correlation of relevant online activity in a region with official reports of outbreaks was measured. Moreover, we conducted trend analysis and compared X and Google Trends in terms of their timeliness and precision in signaling AIV waves. In this study, the effectiveness of utilizing these platforms for outbreak identification is demonstrated, and a comparative analysis is provided between X and Google Trends performance in giving early warning signals for AIV outbreaks.  Furthermore, an ablation study was conducted to assess how each data source contributes to the accuracy of predictions. Specifically, we fitted an auto-regressive time series forecasting model to the reported case numbers, incorporating X and Google Trends data as exogenous variables. Our analysis indicates that both X and Google Trends data independently enhance the accuracy of case number predictions; however, their combined utilization yields even more accurate results.

Despite the increasing interest in leveraging social media and search trends for disease surveillance, their application in animal disease outbreaks, particularly avian influenza, remains significantly underexplored. While prior studies have examined the use of social media and Google Trends data for early identification of human diseases like influenza, our contribution lies in focusing on avian influenza outbreaks—a disease primarily affecting animals. Unlike human-related diseases, which benefit from larger volumes of online data due to broader public engagement, avian influenza surveillance has received limited attention in digital epidemiology. To the best of our knowledge, this is the first study to systematically compare and integrate both social media and Google Trends data for avian influenza outbreak prediction. By addressing this gap, we demonstrate that combining these sources enhances predictive accuracy and provides valuable early warning signals for animal health surveillance. This perspective not only extends the applicability of digital data beyond human health contexts but also highlights its potential for improving early identification strategies for zoonotic diseases. Building on this foundation, our key contributions are as follows:

\begin{itemize}  
\setlength\itemsep{-2pt}  
  \item Highlighting the value of integrating social media and search trends for avian influenza surveillance, demonstrating their complementary roles in enhancing early outbreak identification.
  \item Quantifying the individual and combined contributions of social media and search trends by evaluating their predictive performance in time series models for outbreak forecasting. 
  \item Extending prior research by focusing on avian influenza, a zoonotic disease with lower public engagement, to address the challenges of using digital data for animal disease surveillance.
  \item Demonstrating that integrating multiple digital data sources enhances early warning capabilities for avian influenza outbreaks, even in the context of lower online engagement compared to human diseases. 
  \item Leveraging a fine-tuned LLM to filter irrelevant social media content, demonstrating its effectiveness in improving data quality for avian influenza monitoring compared to traditional machine learning approaches.
\end{itemize}  

\begin{figure}
\begin{center}
\includegraphics[width=1\columnwidth]{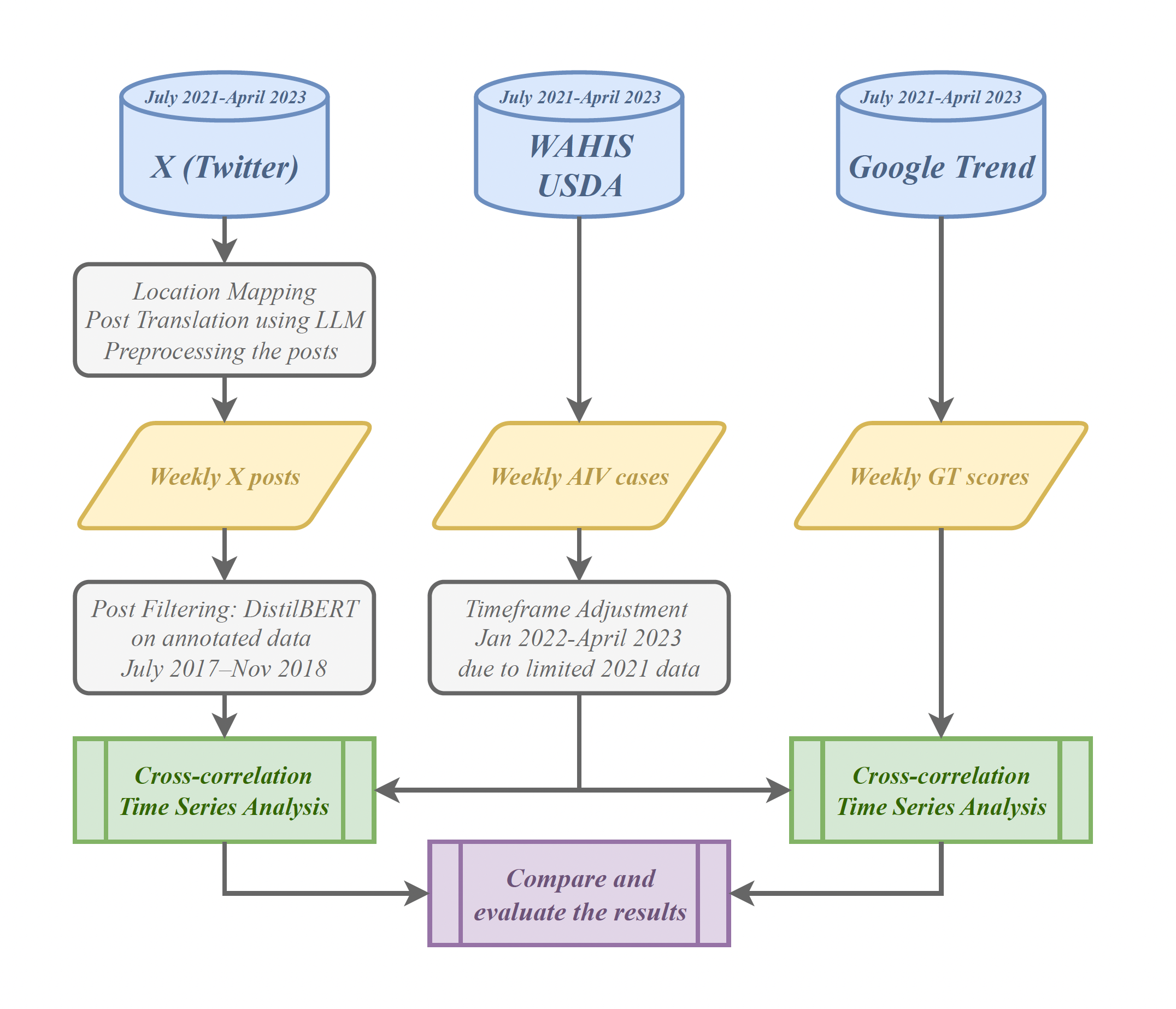}
\vspace{-10pt}
\caption{High-level schematic representation of data collection and analysis}
\vspace{-20pt}
\label{sequence}
\end{center}
\end{figure}

\section{Material and Methods}
\label{method}

In this study, X posts and Google search scores associated with avian influenza outbreaks were collected over 66 weeks, between July 2021 and April 2023. However, upon reviewing the ground truth data for Canada, we observed that there were only three official records of HPAI cases in 2021. To ensure consistency and a more comprehensive analysis, we aligned the timeframe of the study to focus on data from January 2022 to April 2023, during which the data were more representative of outbreak trends. After preprocessing the data, we analyzed the cross-correlation between online activities and officially reported avian influenza cases, representing the number of affected animal units in the outbreak on a weekly basis. This frequency was chosen to mitigate short-term fluctuations in daily data, align with official reporting cycles, and improve the stability for a more reliable analysis of outbreak trends. Following this, the effectiveness of X and Google Trends in predicting early signs of outbreak waves at both national and local levels was investigated and compared. In the present study, an avian influenza outbreak wave denotes a timeframe marked by a notable increase in the number of cases of avian influenza, which is subsequently followed by a decline. Fig.\ref{sequence} illustrates the overall approach of the study.

\subsection{Data and Features}

\label{data}

The selection of X and Google Trends data as primary sources of this study for AIV outbreak identification is based on previous research findings that underscored their efficacy in disease surveillance. In the study conducted by \cite{yousefinaghani2019assessment}, it was found that X, in particular, could be used as a valuable tool for detecting avian influenza outbreaks. Additionally, \cite{lu2018} 's research demonstrated the utility of Google Trends when integrated into a predictive model, in forecasting epidemic avian influenza occurrences. These studies indicated that these digital data sources held significant potential for aiding emergency responders and poultry industries in disease monitoring and early identification. The other reason that encouraged us to use these two sources of data is due to the fact the use of social media and search engines is on the rise according to a study by \cite{pandya2021social}. Also, since 2022 several recent and well-documented AIV outbreaks around the globe have occurred that enabled large-scale data collection \citep{duan2023overview}.

\subsubsection{Ground Truth Data}

\label{truth}

We collected the ground truth data on avian influenza cases reported by two authoritative sources: the World Animal Health Information System (WAHIS) \citep{WAHIS} and the United States Department of Agriculture (USDA) \citep{USDA}. The datasets retrieved from these sources included information such as Event Information, Report Information, Outbreak/Cluster Information, Geographical Area Information, as well as Disease Statistics. The reporting date and computed weekly counts of confirmed avian influenza cases at both the global level and specific country or provincial/state levels were leveraged for the analysis. The United States typically reports HPAI cases exclusively within the "non-poultry including wild birds" category in WAHIS but does not provide data for other species \citep{WAHIS}. According to the notification protocol in WAHIS, when the exact number of cases is unknown or unavailable at the time of reporting, the relevant field is left empty. Consequently, for the analysis of the USA outbreaks, data from both WAHIS and USDA were incorporated and integrated to ensure comprehensive coverage.

\subsubsection{X Data}

\label{X}

To capture real-time discussions related to avian influenza on social media, we retrieved posts through the X Academic Application Programming Interface (API). The API requires us to provide it with a list of keywords, phrases that our target posts on X must contain. Based on prior research, including a study by \cite{yousefinaghani2019assessment}, as well as insights from expert knowledge in the field of avian influenza, we compiled a list of keywords which are illustrated in Fig. \ref{twitter_seed}.In total, around 2M posts were collected from July 2021 to April 2023.

\subsubsection{Google Trends Data}

\label{google}

Google Trends, introduced by Google in May 2006, tracks changes in the popularity of specific search terms in daily, weekly, and monthly intervals. This data is available for specific regions and spans from January 2004. Google Trends normalizes search volumes to show relative popularity compared to overall search activity in the chosen area and time \citep{rech2007discovering, GoogleTrends}. For insights into search patterns on topics related to avian influenza, we utilized Google Trends and focused on the "Interest by Region" feature. This allowed us to identify regions where certain avian influenza-related keywords were most frequently searched. The list of keywords, depicted in Fig. \ref{GT_seed}, was collected through consultation with experts and a thorough review of existing literature. Data collection spanned both the global scale and specific countries, with a particular focus on Canada and the United States for detailed analysis. 

\subsection{Data Pre-processing}

\label{pre-processing}

In the present study, X data preprocessing was conducted through several steps. Initially, X posts were collected using the academic API provided by X, which accepts a list of keywords and returns all the posts containing the queried keywords in the specified time periods and regions if specified. Non-English posts were translated using a large language model. Subsequently, various data-cleaning techniques were applied to remove noise and extraneous information, such as URLs, mentions, hashtags, duplicate posts, and reposts. To improve data quality, NLP techniques were utilized to assess the relevance of each post, as posts could contain unrelated content to avian influenza, such as references to songs, scientific findings, swine influenza, farm sales, and insurance related to bird flu. Filtering out irrelevant posts is an important step in ensuring the accuracy of the outbreak identification process. 

Due to the large size of X data, social media post filtering had to be automated. Therefore, a previously annotated corpus of 4200 randomly selected X posts from July 2017 to November 2018 by domain experts in \cite{yousefinaghani2019assessment} was used to assign "relevant" and "irrelevant" labels to the collected X posts. "Relevant" posts were those related to official reports of avian influenza farm records, emergencies, outbreak consequences, posts and discussions related to AIV outbreaks, and informal reports on individual cases and disease prevention measures. "Irrelevant" posts included jokes, advertisements, and political discussions. This manual labeling resulted in a dataset containing 1,647 "relevant" posts and 2,552 "irrelevant" posts. The DistilBERT model \citep{sanh2019} was fine-tuned using this labeled dataset and applied to automatically label newly collected posts obtained from the X API.

\subsubsection{Translation}

\label{translation}

Certain keywords in the compiled list, such as scientific terms like "H5N1", "H9N2", and "HPAI", appear in some non-English posts, which are inevitably retrieved via the API. To facilitate streamlined analysis, these posts are translated into English using NLLB-200 (facebook/nllb-200-3.3B) \citep{huggingface}, a state-of-the-art LLM developed by Meta AI \citep{team2022}. This process is rather time-consuming and requires high-performance processing infrastructure. We decided to translate the retrieved posts that were in Spanish, Japanese, French, and Italian, the most prominent languages in highly impacted locations, using Compute Canada \citep{Digital} clusters to carry out the process.

\subsubsection{Location Mapping}

\label{mapping}

Several studies have used Geopy, a Python library known for its ability in geolocation tasks, to enhance the precision of user-defined X locations \citep{mounica2022, adams2022, su2021}. In this study, due to the inherent limitations in the accuracy and completeness of manually entered location data on X, Geopy was used in obtaining geographic information, including longitude, latitude, country, and province names. Several posts either lacked profile location information or contained vague location descriptions in the profile field, such as "earth," "worldwide," "everywhere," "nowhere," and "international." Consequently, it was necessary to exclude these posts from further analysis that relied on precise location data. This augmentation of location data was important for subsequent analytical processes, enabling a more geospatially informed analysis of X content concerning avian influenza outbreaks. 

\subsection{Natural Language Processing: DistilBERT}

\label{DistilBERT}

BERT (Bidirectional Encoder Representations from Transformers) is a pre-trained language model developed by Google AI Language \citep{devlin2018bert}. It uses bidirectional processing to understand words in context, unlike traditional models. Through pre-training on a large amount of text data, BERT has been able to capture complex word relationships and create adaptable word representations \citep{han2021pre}. BERT has found applications in chatbots, virtual assistants, search engines, and content recommendation systems. Eq.\ref{eqn:B1}-\ref{eqn:B5} are some common functions used in BERT:

BERT input embeddings:
\begin{equation}
\label{eqn:B1}
\boldsymbol{E} = \text{BERT-Embeddings}(\text{Tokenized-Input})
\end{equation}

Transformer encoder layer:
\begin{equation}
\label{eqn:B2}
\boldsymbol{H}^{l} = \text{Transformer-Encoder-Layer}(\boldsymbol{H}^{l-1})
\end{equation}

BERT output embeddings:
\begin{equation}
\label{eqn:B3}
\boldsymbol{O} = \text{BERT-Output-Embeddings}(\boldsymbol{H}^{L})
\end{equation}

Masked language modeling loss:
\begin{equation}
\label{eqn:B4}
\begin{split}
\mathcal{L}_{\text{MLM}} = & -\sum_{i=1}^{n} \sum_{j=1}^{|\text{Tokenized-Input}_i|} \\
& \log p(\text{Tokenized-Input}_{i,j}|\text{Masked-Input}_{i,j},\theta)
\end{split}
\end{equation}

Next sentence prediction loss:
\begin{equation}
\label{eqn:B5}
\begin{split}
\mathcal{L}_{\text{NSP}} = & -\sum_{i=1}^{n} \log p(\text{IsNext}_i|\text{Output}_i,\theta) \\
& + (1 - \text{IsNext}_i) \log p(\text{NotNext}_i|\text{Output}_i,\theta)
\end{split}
\end{equation}

In the above equations, $\boldsymbol{E}$ represents the input embeddings, $\boldsymbol{H}^{l}$ represents the hidden states at layer $l$, $\boldsymbol{O}$ represents the output embeddings, $\mathcal{L}{\text{MLM}}$ represents the masked language modeling loss, $\mathcal{L}{\text{NSP}}$ represents the next sentence prediction loss, $\text{Tokenized-Input}$ represents the tokenized input sequence, $\text{Masked-Input}$ represents the input sequence with some tokens masked, $\text{Output}$ represents the BERT output sequence, $\theta$ represents the parameters of the BERT model, and $n$ represents the number of training examples.

In this study, the DistilBERT model \citep{sanh2019}, a more computationally efficient variant of the BERT base model \citep{devlin2018bert} was employed. DistilBERT is a method to pre-train a smaller general-purpose language representation model, which can then be fine-tuned with good performance on a wide range of tasks. It retains most of BERT's language understanding capabilities while reducing its size by 40\% and resulting in a higher inference speed. In order to filter out irrelevant content causing noise in the data, we used DistilBERT to appropriately label social media posts relevant to avian influenza outbreak monitoring.

\cite{yousefinaghani2019assessment} utilized a semi-supervised machine learning model in their work with Naive Bayes as the base classifier for the automatic labeling of social media posts. To train the model, a manually annotated dataset was used to determine the label of unlabeled social media posts. In this study, we utilized the same dataset provided by \cite{yousefinaghani2019assessment}, where "relevant" and "irrelevant" labels were manually assigned to approximately 4,200 sample X posts from July 2017 to November 2018. These posts were randomly selected across monthly periods, and guidelines were defined to assist experts in annotating posts. The manually annotated dataset included 1,647 positive labels (indicating relevance) and 2,552 negative labels (indicating irrelevance).

In order to train the model, the manually labeled dataset was partitioned into two subsets for training and evaluation. After fine-tuning DistilBERT on the training subset, it was able to achieve an 89.5\% accuracy score on the previously unseen evaluation subset. Notably, DistilBERT outperformed the Naive Bayes semi-supervised learning model used in \cite{yousefinaghani2019assessment}'s study with an average accuracy of 78.4\%. 
Following the fine-tuning process, the DistilBERT model was then applied to classify new posts collected between January 2022 and April 2023.  {After applying the classification model, further filtering was conducted to exclude posts without location information, resulting in a refined dataset of approximately 210K relevant posts that spanned over 66 weeks.}

\subsection{Cross-correlation Analysis}
\label{models}

In this study, our primary objective was to assess the potential of X and Google Trends data for early warning of avian influenza outbreaks. Specifically, this study aimed to achieve two key sub-objectives: (1) to examine the correlation between X and Google Trends data with official reports of avian influenza cases, and (2) to test the effectiveness of X and Google Trends as tools for early warning of potential avian influenza outbreaks. 

Cross-correlation is a method for assessing the similarity between two waveforms. This method was used to assess the correlation between case numbers and post count, or Google Trends score, considering the displacement or time lag between these variables. The specific lag at which the two variables exhibited the highest degree of correlation was determined by calculating the cross-correlation. This lag value, representing the maximum coefficient, is the one reported in the tables. Pearson correlation coefficient was utilized to measure the degree of correlation between time series data representing activities on these platforms within each province/state or globally and the corresponding reported avian influenza cases. Correlation coefficients, ranging from 0 to 1, measure the relationship between two data series. A coefficient of 1 signifies a perfect match, while 0 indicates no noticeable correlation.

The cross-correlation between discrete functions \(f\) and \(g\) at shift \(n\) is defined as:

 {
\begin{equation}
\label{eqn:cc}
\mathrm{CC} = (f \star g)[n] \overset{\text{def}}{=} \sum_{m=-\infty}^{\infty} \overline{f[m]} g[m+n]
\end{equation}
}

This equation quantifies the relationship between two discrete signals \(f\) and \(g\) by considering their complex conjugates and time-shifted products \citep{lyon2010discrete}.

A requirement of cross-correlation analysis is that the data should be stationary. Stationarity refers to a time series with a constant mean and variance over time, without a persistent trend. To assess stationarity, both the Augmented Dickey-Fuller (ADF) and Kwiatkowski-Phillips-Schmidt-Shin (KPSS) tests are commonly applied. If the null hypothesis of the ADF test is rejected, it suggests that the time series is stationary, making it suitable for further analysis. Conversely, failing to reject the null hypothesis of the KPSS test indicates that the series is stationary. If either test suggests that the data is not stationary, applying integration may be necessary to achieve stationarity for time series analysis. This sequence of tests ensures that the cross-correlation results are reliable and not influenced by trends or non-stationary patterns in the data.

\subsection{ {Time Series Forecasting}}

Time series forecasting is a well-explored problem in machine learning due to its wide range of applications, including those in epidemiology. While modern deep learning methods, such as recurrent neural networks and transformer architectures, have garnered attention for modeling complex patterns in large datasets, they often demand substantial data and computational resources. In contrast, classical approaches, such as state-space models and information filters, are more practical and effective for scenarios with limited data. These methods are not only easier to interpret but also grounded in strong theoretical foundations, which are crucial for designing reliable decision support systems. In this study, constrained by the limited size of our dataset, we adopted a classical approach: Auto-Regressive (AR) prediction methods \citep{durbin2012time}, which remain highly relevant in time series forecasting. These methods predict future values based on relationships with past observations. Among them, the Seasonal Auto-Regressive Integrated Moving Average eXogenous (SARIMAX) model \citep{box2015time} stands out for its flexibility and effectiveness, making it an ideal choice for our data.

The SARIMAX model builds upon the basic AR framework by incorporating several components. A basic AR (AutoRegressive) model uses a linear combination of past observations to predict the value of a time series at a point based on the past $p$ observations. In order to address non-stationary time series, sequences where statistical properties such as mean and variance change over time, the model would \textbf{I}ntegrate past $d$ values by subtracting them from their preceding values, effectively transforming the series into a stationary form. Similarly, the \textbf{M}oving \textbf{A}verage component models the influence of past forecast errors over $q$ lags, capturing short-term fluctuations. Moreover, time series such as ours exhibit seasonal patterns, where observations repeat at regular intervals. To handle these, the model includes \textbf{S}easonal counterparts to these components over a period of $s$ observations, controlled by three additional hyperparameters, $P$, $Q$, and $D$. Finally, the model can incorporate e\textbf{X}ogenous variables, variables external to the time series itself but potentially influencing its behavior. These exogenous variables provide valuable context to the intrinsic patterns of the time series, enabling the model to account for external factors that drive variation. By integrating these external influences, SARIMAX can generate more accurate and meaningful forecasts, particularly in scenarios where external indicators are expected to correlate strongly with the observed patterns.

\section{Results}

\subsection{Global Seed Word Analysis}

\begin{figure}[h!]
\vspace{-2mm}
    \begin{center}
    \includegraphics[width=0.9\columnwidth]{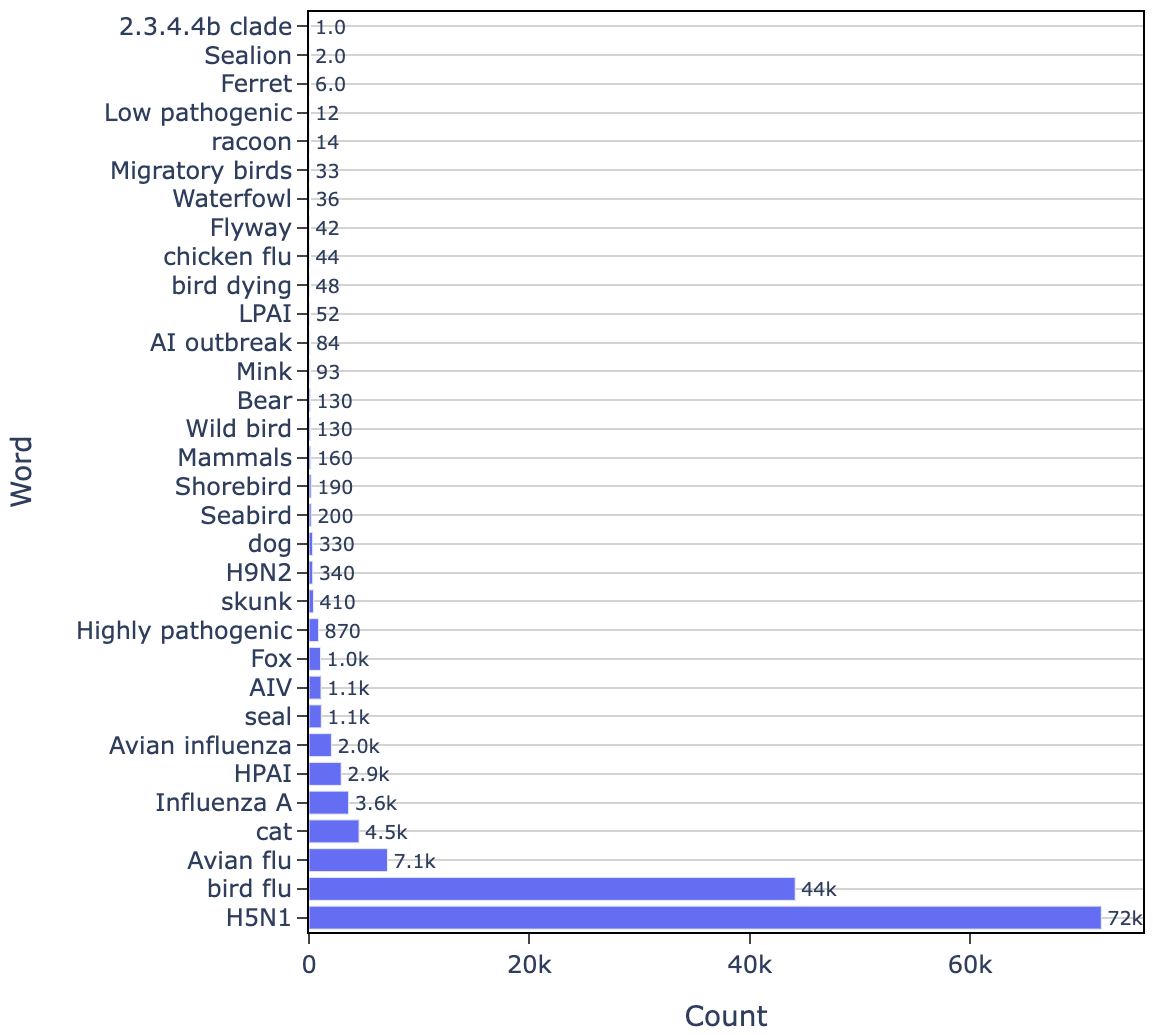}
    \vspace{-2mm}
    \caption{X Seed Word Bar Chart}
    \vspace{-20pt}
    \label{twitter_seed}
    \end{center}
\end{figure}

\begin{figure}[h!]
\vspace{-2mm}
\begin{center}
\includegraphics[width=0.9\columnwidth]{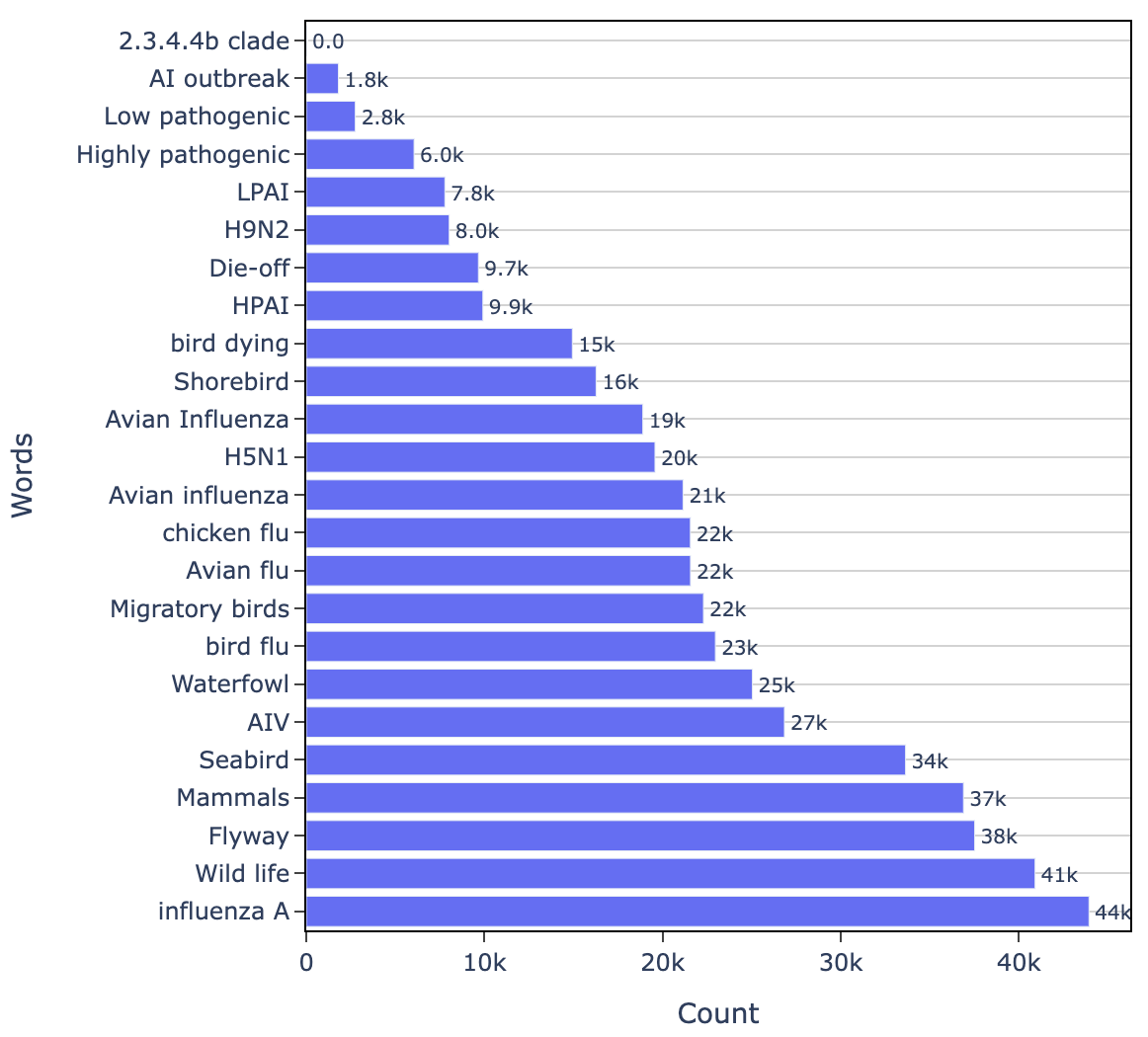}
\vspace{-2mm}
\caption{Google Trends Seed Word Bar Chart}
\vspace{-20pt}
\label{GT_seed}
\end{center}
\end{figure}

Seed words, which refer to a curated list of keywords utilized for the retrieval of posts via the X academic API, were carefully chosen using prior studies by \cite{yousefinaghani2019assessment}, an analysis of initial post collections for the identification of additional relevant keywords, and input from subject matter experts. These seed words were employed for the global-level data collection using X and Google Trends data across various languages. The frequency of these seed words in the collected posts is presented in Fig. \ref{twitter_seed}, revealing 'H5N1' and 'bird flu' as the most frequently mentioned terms, while 'sealion' and '2.3.4.4b clade' were among the least common. An interesting observation was the frequent use of the term "cat" in these posts, possibly due to recent instances of avian influenza transmission to mammals, underscoring the evolving nature of the disease and global concerns. Although 'wild bird' and 'migrating bird' were expected to be more frequent, given the impact of wild birds on the spread of AIV outbreaks, they displayed lower frequencies on X. This could be due to the fact that X posts are generally by the public rather than experts. The figures collectively demonstrate the importance of 'H5N1' and 'bird flu' in discussions about avian influenza on X. Furthermore, Fig. \ref{GT_seed} illustrates the search counts of seed words in Google Trends, with 'Influenza A,' 'Wildlife,' 'Flyway,' 'Mammals,' and 'Seabird' were the most frequently searched. 

\subsection{Keyword Analysis}

\begin{figure}[h!]
\vspace{-2mm}
\begin{center}
\includegraphics[width=0.9\columnwidth]{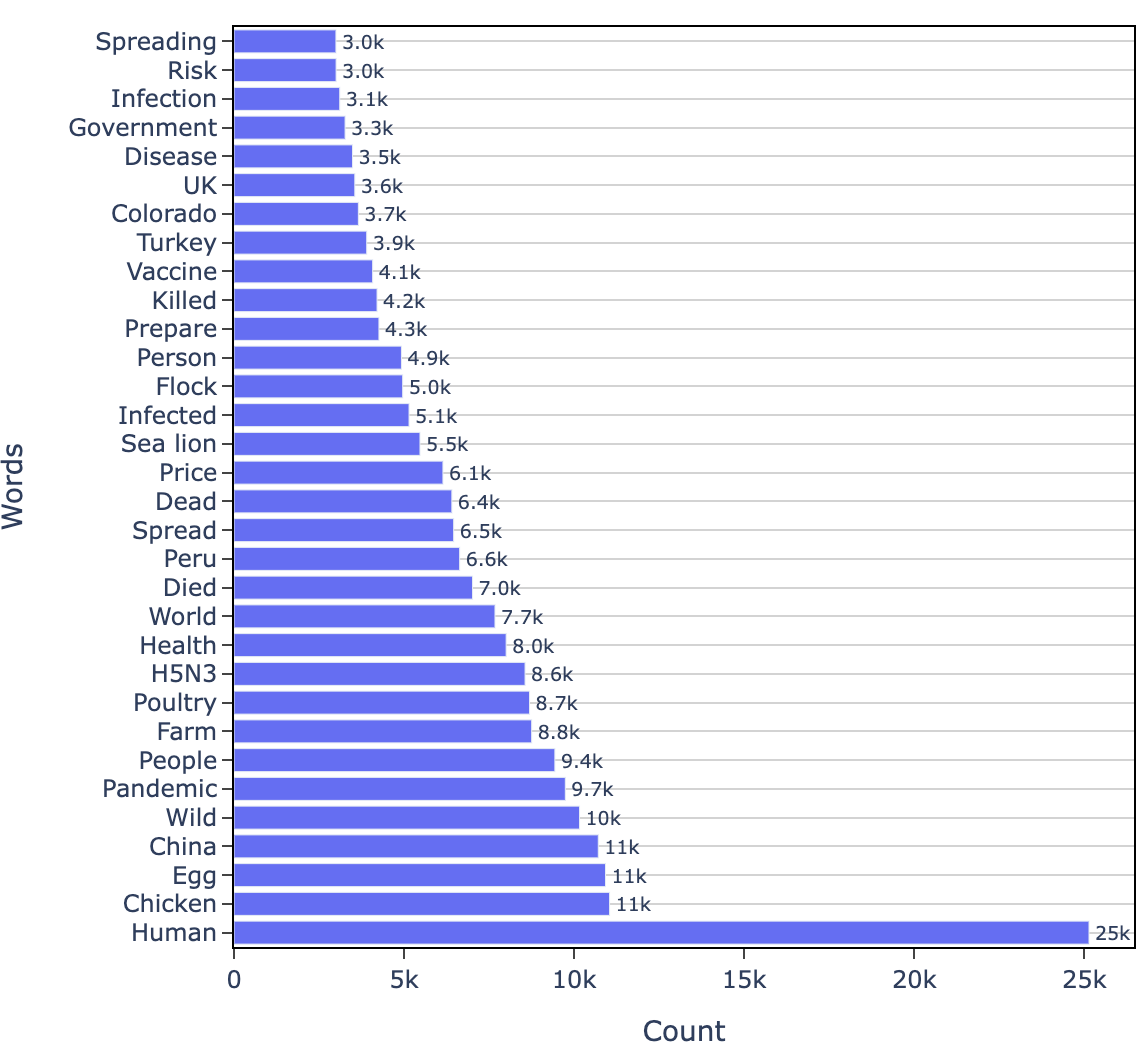}
\vspace{-2mm}
\caption{X Frequent Keywords Bar Chart}
\vspace{-20pt}
\label{keyword}
\end{center}
\end{figure}

Fig.\ref{keyword} illustrates a bar chart with the most frequent words found in relevant X posts, excluding the seed words. The figure demonstrates some of the prevailing discourse surrounding avian influenza. The presence of terms such as "human," "People," and "Person" among frequently used keywords suggests heightened public concern over sporadic human cases associated with poultry exposure during the ongoing outbreak of H5N1 viruses. Certain geographical terms were also observed on the chart, namely 'UK' and 'Colorado', both of which are hotbeds with frequent AIV reported cases based on WAHIS and USDA for avian influenza. Another frequent term that shows up on the chart is 'Egg'. This could be linked to the fact that egg prices often surge during avian influenza outbreaks. During the period of collected historical data, egg prices soared from 2\$ to 6\$ cents per dozen between January and December 2022 which revealed substantial economic implications for the egg industry \citep{usdaAvianInfluenza}.

\subsection{Map Distribution}

\begin{figure*}
    \centering
    \includegraphics[width=1.8\columnwidth]{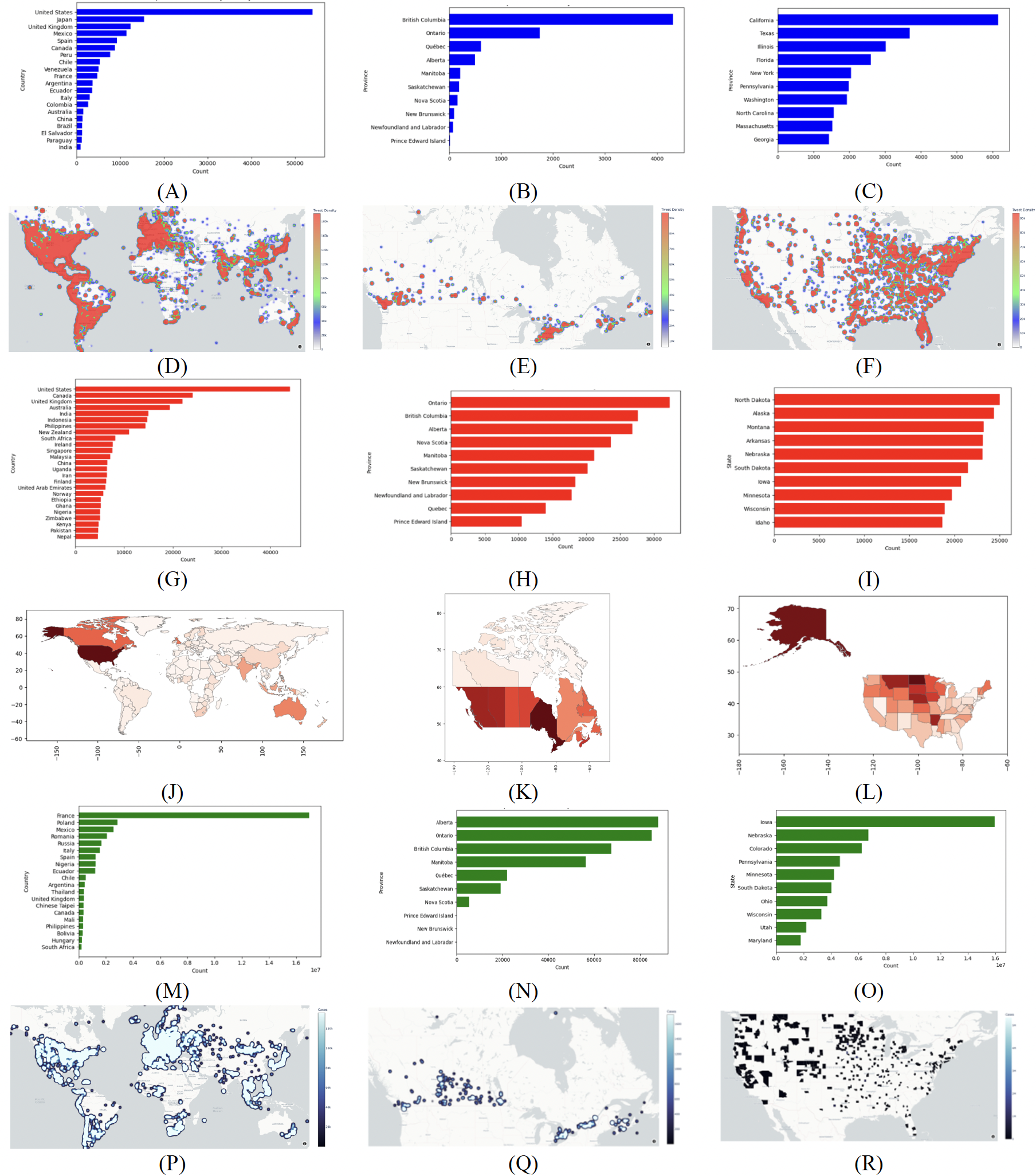}
    \caption{X posts frequencies (A, B, C), heatmaps of social media activity (D, E, F), Google Trends score (G, H, I), heatmaps of search scores (J, K, L), avian influenza case density (M, N, O), and heatmap of avian influenza reported cases (P, Q, R). The figures demonstrate activity on a global scale (A, D, G, J, M, P), in Canada (B, E, H, K, N, Q), and the USA (C, F, I, L, O, R).} 
    \vspace{-10pt}
    \label{map}
\end{figure*}

In Fig. \ref{map}\textcolor{cyan}{A,D}, a heatmap and bar chart are presented that illustrate global X activity on avian influenza. The data shows that the USA leads in X engagement with over $53K$ posts, followed by Japan, the United Kingdom, and Mexico in Fig. \ref{map}\textcolor{cyan}{A}. This observation shows the USA's substantial contribution to X conversations. In Canada, British Columbia stands out with the highest post count among provinces, indicating stronger social media activities in this region as shown in Fig. \ref{map}\textcolor{cyan}{B}. The state-based analysis of post counts depicted in Fig. \ref{map}\textcolor{cyan}{C} shows California, Texas, and Illinois are the top three locations with the most frequent post counts on X. It is also worth mentioning that, due to missing or invalid geolocations in X data, many posts could not be associated with an accurate location. Therefore, public engagement in various locations could be underestimated. The heatmap depicted in Fig. \ref{map}\textcolor{cyan}{D} highlights increased public engagement in X, particularly in the northern and southern parts of both North America and Europe.

The heatmap and the bar chart of Google Trends scores related to avian influenza on a global level, in Canada, and the USA are also depicted in Fig. \ref{map}. The numbers in Fig. \ref{map}\textcolor{cyan}{G} highlight the interest and engagement from USA and Canada users that indicate a notable information-seeking behavior regarding avian influenza. Ontario, BC, and Alberta are the top 3 provinces in Canada, with high Google Trends scores that could be a sign of potential local outbreaks. Among the states with high Google Trends scores, North Dakota, Alaska, and Montana are the leading three locations in the USA as shown in Fig. \ref{map}\textcolor{cyan}{I}. The highest Google Trends scores as visualized in Fig. \ref{map}\textcolor{cyan}{J} are more concentrated within the northern parts of North America and Australia. Additionally, it's worth noting that Google usage is restricted in some countries such as China. Therefore, this limitation could potentially affect the representation of interest and engagement levels in those regions.

Fig. \ref{map} offers a map and bar chart showing avian influenza case density on a global scale, in Canada, and in the USA. France leads globally in reported cases, followed by Poland and Mexico. Notably, France maintains transparent communication with importing countries to ensure necessary information exchange \citep{remongin}. Therefore, it is the leading country in Fig. \ref{map}\textcolor{cyan}{M} among other countries. In contrast, the USA's reporting process to WOAH is different, and our analysis had to rely on USDA data for the USA related analysis. Among the provinces in Canada with the highest number of avian influenza reported cases, Alberta, Ontario, and BC are the top 3 locations with frequent cases of this virus as depicted in Figures. \ref{map}\textcolor{cyan}{N,Q}. As can be seen in Figures. \ref{map}\textcolor{cyan}{O,R}, Iowa, Nebraska, and Colorado are also the top three states with the highest avian influenza reported case density.

Fig. \ref{map} provides a comprehensive view of global, Canadian, and USA trends in social media activity, online searches, and avian influenza outbreaks. Comparing X data, Google Trends scores and avian influenza cases reveals interesting patterns. While X and Google Trends offer insights into online engagement and information-seeking behavior, they may not always directly be aligned with actual disease prevalence. 

\subsection{Visual Trends}

\begin{figure*}
\begin{center}
\includegraphics[width=1.8\columnwidth]{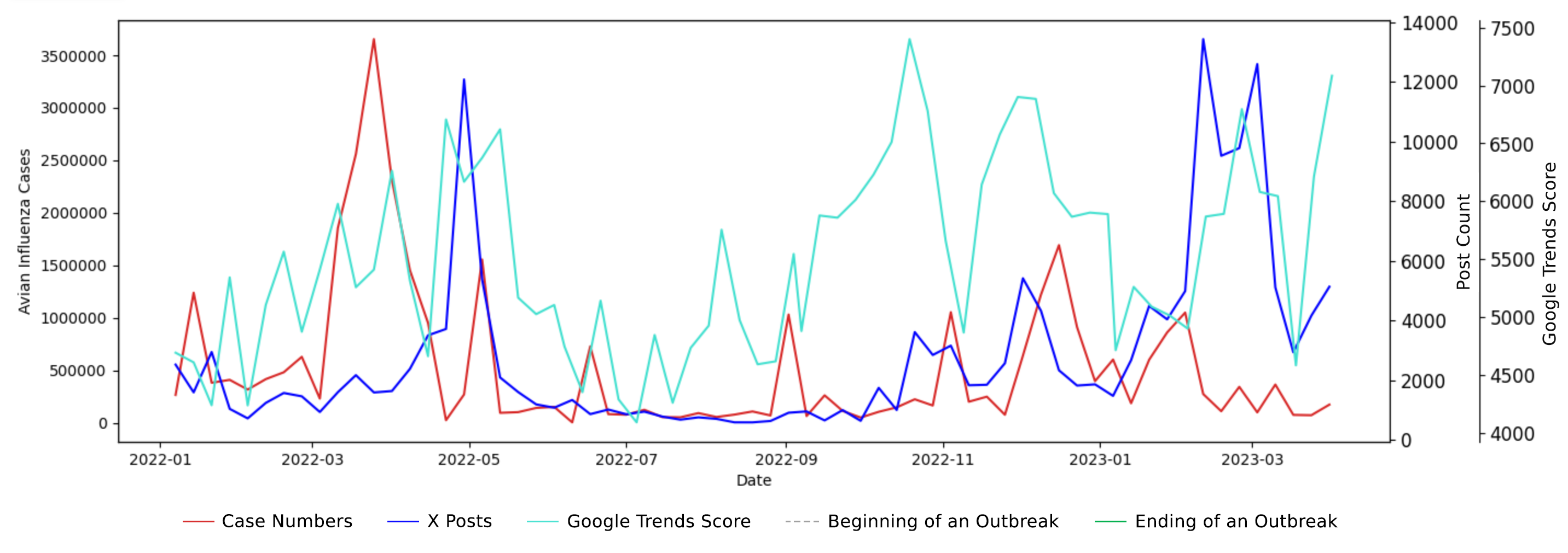}
\vspace{-5pt}
\caption{Weekly Global Trend of Post Count, Google Trends Score, and Avian Influenza Cases. Specific timeframes for outbreak waves cannot be depicted on a global scale.}
\vspace{-20pt}
\label{Global_trend}
\end{center}
\end{figure*}

\begin{figure*}
\begin{center}
\includegraphics[width=1.8\columnwidth]{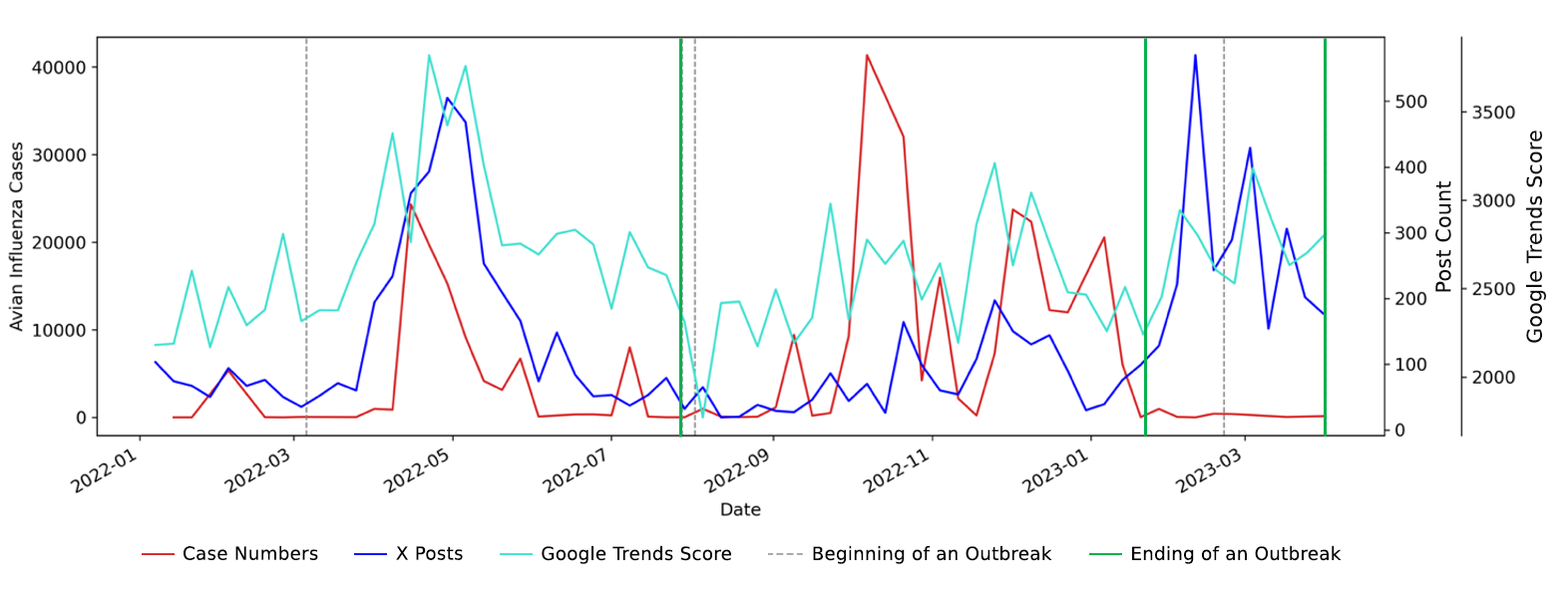}
\vspace{-5pt}
\caption{Weekly Trend of Post Count, Google Trends Score, and Avian Influenza Cases in Canada. Depicted outbreak waves were reported by the Canadian Food Inspection Agency (CFIA). The gray dashed line represents 3 weeks prior to the outbreak start date, and the green line shows the end of the wave.}
\vspace{-20pt}
\label{Canada_trend}
\end{center}
\end{figure*}

\begin{figure*}
\begin{center}
\vspace{-8pt}
\includegraphics[width=1.8\columnwidth]{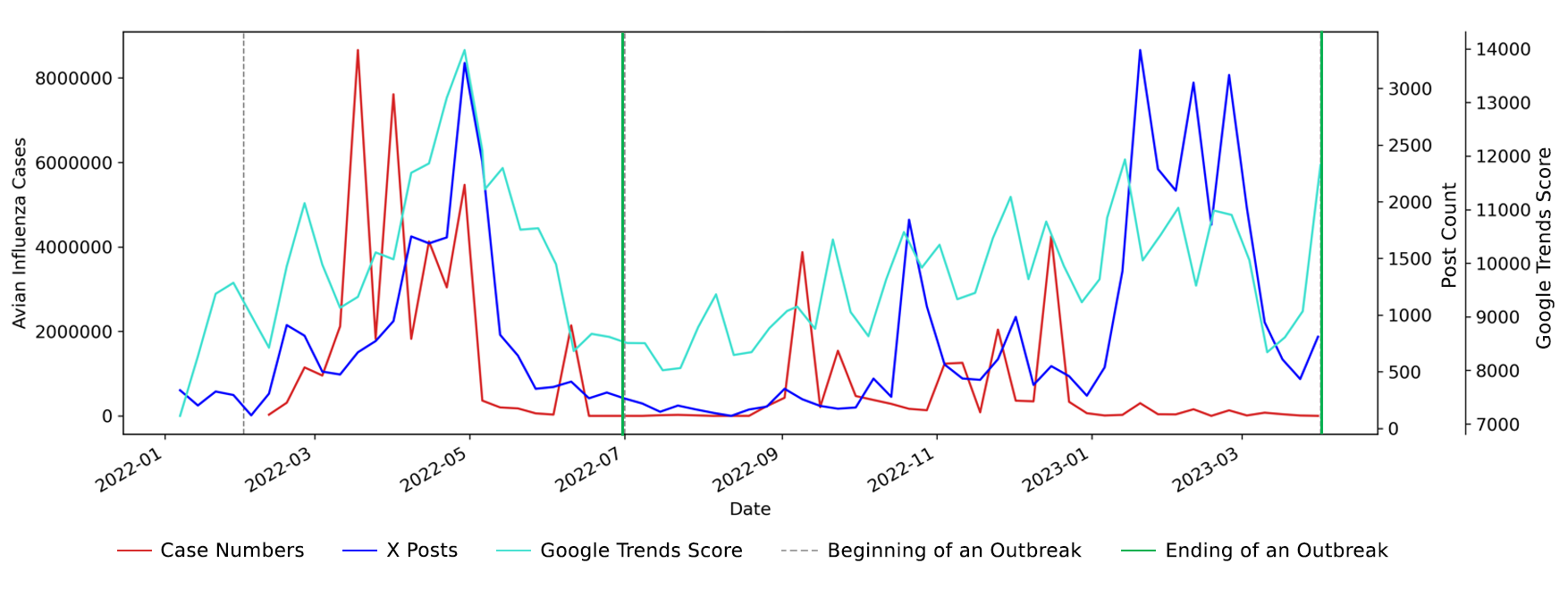}
\vspace{-5pt}
\caption{Weekly Trend of Post Count, Google Trends Score, and Avian Influenza Cases in the USA. Depicted outbreak waves were reported by the USDA. The gray dashed line represents 3 weeks prior to the outbreak start date, and the green line shows the end of the wave.}
\vspace{-10pt}
\label{USA_trend}
\end{center}
\end{figure*}

In this section, weekly data on X activity and Google Trends scores related to avian influenza keywords were visualized alongside the reported number of avian influenza cases, both globally and regionally, including Canada (Ontario, Alberta, and BC) and the USA (Iowa, Nebraska, Colorado). The selected provinces/states are based on the most frequent cases reported by WAHIS and USDA. In charts \ref{Global_trend}-\ref{USA_trend}, the number of relevant posts on X is visualized alongside Google Trends scores, as well as officially reported cases. Fig \ref{Global_trend} encompasses all the collected data for global-scale analysis, while Figures \ref{Canada_trend} and \ref{USA_trend} focus on Canada and the USA, respectively.

The data suggest that X activity in most states and provinces exhibited slightly earlier peaks compared to Google. This pattern can also be seen in the visual trends in the Supplementary Material. However, when considering trends during the second wave of outbreaks in Canada and The USA, Google searches displayed more prominent peaks. Generally, as the number of cases began to decline after peak periods, a gradual reduction in both X posting and Google searches was noted. This decline may be attributed to knowledge saturation, wherein the public and practitioners' interest and engagement decrease as they become more informed about the outbreak. It is noteworthy to mention that avian influenza-related terms were more frequently searched on Google across all geographical locations.

\label{result}
\subsection{Correlation Analysis}

Prior to performing the cross-correlation analysis, we verified the stationarity of the datasets using both the ADF and KPSS tests to provide a comprehensive assessment. Both tests were performed with a constant term, as initial data patterns suggested this was the most suitable configuration. The results indicated that all datasets, except for BC in the ADF test, were stationary ($p$-values $<0.05$). To further verify stationarity, we also conducted the KPSS test, which confirmed stationarity across all datasets. For the BC dataset, where only the ADF test did not confirm stationarity, first-order differencing was applied, resulting in the data achieving stationarity. This ensured that the stationarity requirement for cross-correlation analysis was met across all cases.

Table~\ref{tab:global_cross} summarizes the global temporal correlations from January 2022 until April 2023 between X and Google Trends activity and avian influenza reported cases. Google Trends exhibited a statistically significant ($p$-value $=0.01$) positive correlation ($r=0.27$) with avian influenza cases, with a $-2$ weeks lag. This result is due to the growth in online activities on this platform that preceded peaks in reported avian influenza cases. Furthermore, the cross-correlation of global X data with the reported case numbers showed a positive correlation ($r=0.25$) with a $-1$ week lag and a $p$-value of $0.0509$, which is only slightly higher than the significance level of 0.05. The absence of a significant correlation in social media on a global scale may be attributed to variations in X usage worldwide, including regions with limited access to social media. 

\begin{figure*}
\begin{center}
\vspace{-5pt}
\includegraphics[width=2\columnwidth]{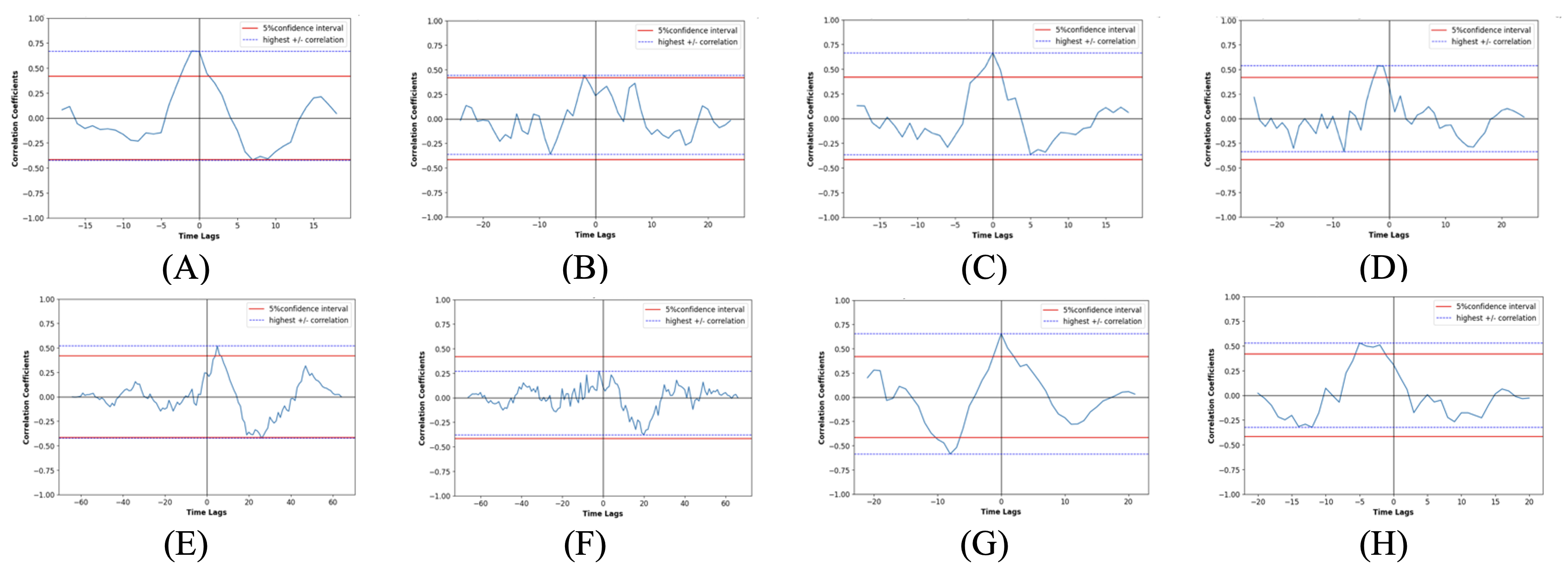}
\caption{Statistically significant cross-correlation trends with different lags for X (A, B, E, G, H) and Google Trends (C, D, F), on a global scale (E, F), in Canada (A, B, C, D), in the USA (G), and in BC (H). The horizontal red lines in the graph denote the 95\% confidence interval for the correlation coefficients. Additionally, the dotted blue lines represent the maximum and minimum values of the correlation coefficients.}
\vspace{-10pt}
\label{lags}
\end{center}
\end{figure*}

Fig. \ref{lags}\textcolor{cyan}{E,F} visually illustrates correlation coefficients across a wide range of lag values on a global level. As can be seen in Fig. \ref{lags}\textcolor{cyan}{F} the correlation coefficient values are inside the 95\% confidence interval denoted by horizontal red lines. This suggests that the observed correlation coefficients are likely to be statistically significant, as they are within the expected range based on the confidence interval. Moreover, in this study, lags exceeding $-4$ weeks are not reported in the analysis, as they may not yield relevant information for timely outbreak identification. 

\begin{table*}[ht]
\centering
\vspace{-10pt}
\caption{Global Cross-Correlation Results}\label{tab:global_cross}
\begin{tabular}{ccccccc}
\toprule
\textbf{Location} & \textbf{Media} & \textbf{Cases} & \textbf{Timeframe}  & \textbf{Lag \footnotesize(Weeks)} & \textbf{Coeff.} & \textbf{ {P-value}} \\
\midrule
Global & X & 34M & 2022-01-01 to 2023-04-01 & -1 & 0.25 & 0.0509 \\
Global & Google Trends & 34M & 2022-01-01 to 2023-04-01 & -2 & 0.27 & 0.0199 \\
\bottomrule
\end{tabular}
\end{table*}

\begin{table*}[ht]
\centering
\vspace{-15pt}
\caption{Statistically Significant Cross-Correlation Results}\label{tab:stat_cross}
\begin{tabular}{ccccccc}
\toprule
\textbf{Location} & \textbf{Media} & \textbf{Cases} & \textbf{Timeframe}  & \textbf{Lag  \textsubscript{Weeks}} & \textbf{Coeff.} & \textbf{ {P-value}} \\
\midrule
Global & Google Trends & 34M & 2022-01-01 to 2023-04-01 & -2 & 0.27 & 0.0199 \\
Canada (1st Wave) & X & 93K & 2022-03-06 to 2022-07-28 & -1 & 0.67 & 0.0229 \\
Canada (2nd Wave) & X & 222K & 2022-08-02 to 2023-01-22 & -2 & 0.44 & 0.0269 \\
Canada (1st Wave) & Google Trends & 93K & 2022-03-06 to 2022-07-28 & -1 & 0.52 & 0.0409 \\
Canada (2nd Wave) & Google Trends & 222K & 2022-08-02 to 2023-01-22 & -2 & 0.54 & 0.0179 \\
BC (2nd Wave) & X & 58K & 2022-08-22 to 2023-01-22 & -2 &  {0.64} &  {0.0129} \\
USA (1st Wave) & X & 40M & 2022-02-01 to 2022-07-01 & -1 & 0.46 & 0.0339 \\
\bottomrule
\end{tabular}
\end{table*}

Table~\ref{tab:stat_cross} illustrates an overview of the significant cross-correlation findings. The results demonstrate the temporal correlations between online activity and avian influenza cases across different locations and timeframes. Utilizing data from X and Google Trends, the results highlight the potential of these platforms as early indicators of avian influenza outbreaks. Notably, all the correlations presented in Table~\ref{tab:stat_cross} are positive which shows a positive link between online activity and avian influenza cases. The lag period reported for each location reflects the time point at which online activity aligns with avian influenza case reports. This indicates the period where online activity closely precedes an increase in reported avian influenza cases and demonstrates the capability to predict the outbreak 1-2 weeks in advance.

As previously mentioned, the findings indicated a statistically significant positive correlation ($r=0.27$) between Google Trends data and reported avian influenza cases that is able to predict outbreaks 2 weeks earlier. Positive correlations were also observed in specific regions such as Canada (across different waves), British Columbia (BC), and the USA (in the first wave), with corresponding lag periods. The positive correlation between online activities and avian influenza outbreak waves shows the potential of social media data and search engines as valuable tools for early identification and surveillance of avian influenza outbreaks. However, the absence of statistical significance in other waves/locations such as Ontario, Alberta, and the selected USA states may be attributed to factors such as sample size limitations, data quality, the complexity of avian influenza outbreaks, and the presence of confounding variables, which can influence the ability to establish statistically significant correlations. Fig. \ref{lags} provides a visual representation of these statistically significant cross-correlation trends with different lags. As illustrated in Fig. \ref{lags}, the majority of values fall within the 95\% confidence interval, marked by the red lines. Fig. \ref{lags} also shows that the majority of reported correlation coefficients with a negative lag value are among the highest values, as indicated by the dashed horizontal blue line.

\subsection{Correlation Analysis: Canada}

Cross-correlation analysis of avian influenza outbreaks in Canada shows interesting patterns across different waves and online platforms, as shown in Table~\ref{tab:canada_cross}. The timeframes of the avian influenza outbreak waves by CFIA are reported in Table~\ref{tab:canada_cross}. The first wave occurred from March 2022 to July 2022, followed by the second wave between August 2022 and January 2023, and the third wave from February 2023 to April 2023. In the first and second waves, both X and Google Trends demonstrated statistically significant ($p$-value $<0.05$) positive correlations with avian influenza reported cases, with respective lags of $-1$ and $-2$ weeks. Notably, for the 3rd wave in Canada.  {The cross-correlation analysis for Canada's third wave with both X and Google Trends yielded high $p$-values, indicating no significant linear relationship. This outcome is likely due to the exceptionally low case counts during this wave and does not rule out potential non-linear patterns.} Overall, the results demonstrate a stronger correlation between the prevalence of avian influenza cases and the level of online activity in the first wave, with correlation coefficient values of 0.67 and 0.52 for X and Google Trends, respectively. However, during the third wave, while correlations remained positive ($r(X)=0.32, r(Google Trends)=0.36$), the results were not statistically significant ($p$-value $>0.05$). This could be due to the lack of data from X and Google Trends. Lag periods illustrated in Fig. \ref{lags}\textcolor{cyan}{A,B} for X and Fig. \ref{lags}\textcolor{cyan}{C,D} for Google Trends implied that increased online activity may precede rises in reported cases.

Furthermore, analysis of Ontario, Alberta, and British Columbia data during the distinct waves of the avian influenza outbreak revealed varying correlations between X and Google Trends data with reported cases. Although suggestive correlations were observed, they did not reach statistical significance, which indicates a potential but not conclusive correlation. Detailed correlation results for each province along with corresponding lag periods can be found in the Supplementary Material tables. Additionally, during the selected timeframe of this study, most Canadian provinces did not experience a significant third wave of avian influenza. Therefore, the results showed a weak and statistically insignificant correlation between online activity and reported cases. This research highlights the potential of social media data, particularly in the early waves of avian influenza outbreaks, as valuable tools for surveillance and early warning, as depicted in the first and second waves in Fig. \ref{lags}\textcolor{cyan}{A,B,C,D}. 

\begin{table*}[ht]
\centering
\vspace{-10pt}
\caption{Canada Cross-Correlation Results}\label{tab:canada_cross}
\begin{tabular}{ccccccc}
\toprule
\textbf{Location} & \textbf{Media} & \textbf{Cases} & \textbf{Timeframe}  & \textbf{Lag  \textsubscript{Weeks}} & \textbf{Coeff.} & \textbf{ {P-value}} \\
\midrule
Canada (1st Wave) & X & 93K & 2022-03-06 to 2022-07-28 & -1 & 0.67 & 0.0229 \\
Canada (2nd Wave) & X & 222K & 2022-08-02 to 2023-01-22 & -2 & 0.44 & 0.0269 \\
Canada (3rd Wave) & X & 746 & 2023-02-21 to 2023-04-01 & -1 & 0.32 & 0.7882 \\
Canada (1st Wave) & Google Trends & 93K & 2022-03-06 to 2022-07-28 & -1 & 0.52 & 0.0409 \\
Canada (2nd Wave) & Google Trends & 222K & 2022-08-02 to 2023-01-22 & -2 & 0.54 & 0.0179 \\
Canada (3rd Wave) & Google Trends & 746 & 2023-02-21 to 2023-04-01 & -3 & 0.36 & 0.8201 \\
\bottomrule
\end{tabular}
\end{table*}

\subsection{Correlation Analysis: USA}

Table~\ref{tab:usa_cross} summarizes the cross-correlation results for the avian influenza outbreak in the United States across two waves reported by USDA. The first wave spanned from February 2022 to July 2022, followed by the second wave from July 2022 to April 2023. During the first wave, X data and avian influenza case numbers displayed a positive correlation coefficient ($r=0.46$) with a $-1$ week lag. However, in the second wave, this correlation was weak and statistically insignificant. Similarly, Google Trends data showed positive correlation values of 0.39 and 0.27 in both waves, respectively; however, these values were not statistically significant. 

Among the three states with the highest avian influenza case frequencies, Iowa, Nebraska, and Colorado, positive correlations were observed in both X and Google Trends data. However, the state-wise cross-correlation results reported in the Supplementary Material tables were not statistically significant ($p$-value $>0.05$). This may be due to limited data availability and the complexity of these trends. While not statistically significant, these findings highlight the potential link between online activity and avian influenza cases, emphasizing the need for further research and data analysis. Detailed cross-correlation results for each state, along with corresponding lag periods, can be found in the Supplementary Material.

\vspace{-10pt}
\begin{table*}[ht]
\centering
\caption{USA Cross-Correlation Results}\label{tab:usa_cross}
\begin{tabular}{ccccccc}
\toprule
\textbf{Location} & \textbf{Media} & \textbf{Cases} & \textbf{Timeframe}  & \textbf{Lag  \textsubscript{Weeks}} & \textbf{Coeff.} & \textbf{ {P-value}} \\
\midrule
USA (1st Wave) & X & 40M & 2022-02-01 to 2022-07-01 & -1 & 0.46 & 0.0339 \\
USA (2nd Wave) & X & 18M & 2022-07-01 to 2023-04-01 & -1 & 0.11 & 0.4835 \\
USA (1st Wave) & Google Trends & 40M & 2022-02-01 to 2022-07-01 & -1 & 0.39 & 0.0739 \\
USA (2nd Wave) & Google Trends & 18M & 2022-07-01 to 2023-04-01 & -1 & 0.27 & 0.0809 \\
\bottomrule
\end{tabular}
\end{table*}

\subsection{Ablation Study}

To analyze the effect and contribution of online activity on X and Google Trends score in predicting AIV outbreaks, we performed an ablation study, where we examine the impact of using these online outlets as exogenous variables in a SARIMAX model. For this analysis, we solely focused on global data due to the limited weekly datasets available for individual outbreaks in Canada and the USA. By leveraging the global data, we demonstrated the value of incorporating exogenous variables like online activity metrics for outbreak prediction while acknowledging the challenges of scale in localized studies.

For this analysis, the data is split into subsets, for training and evaluation. Using an implementation of SARIMAX found in the Stats-Models Python library, we performed a grid search to find the optimal hyperparameters for the model. We fitted the SARIMAX model to the WAHIS case numbers under four different scenarios; (i) without using any exogenous variables, (ii) only using X posts count as an exogenous variable, (iii) only using Google Trends scores as an exogenous variable, and (iv) using both as exogenous variables.  {To ensure the validity of our model, we checked for multicollinearity between X posts and Google Trends. The Pearson correlation coefficient (0.20) and Variance Inflation Factor ($VIF = 1.04$) indicate low correlation, confirming that both variables contribute complementary information.} To assess the performance of the model under these scenarios, we ran each experiment 30 times with randomly initialized weights, and calculated their average $\mathrm{R}^2$ score. $\mathrm{R}^2$ is a scale-independent regression metric that ranges between $(-\infty,1]$.

\begin{table}[]
\caption{ {SARIMAX Performance on Global Data with and without Lagged Exogenous Variables}}
\label{tab:ablation}
\begin{tabular}{@{}cccc@{}}
\toprule
\textbf{Scenario} & \textbf{X Posts} & \textbf{Google Trends} & \textbf{\boldmath{$\mathrm{R}^2$} Score} \\
\midrule
\multicolumn{4}{c}{\small\textit{ {Exog. without lag (Train: 52w, Eval: 13w)}}} \\
\midrule
(i)  &            &             & $0.2032 \pm 0.0031$ \\
(ii) & \checkmark &             & $0.2219 \pm 0.0035$ \\
(iii)&            & \checkmark  & $0.2270 \pm 0.0038$ \\
(iv) & \checkmark & \checkmark  & $0.2491 \pm 0.0027$ \\   
\midrule
\multicolumn{4}{c}{\small\textit{ {Exog. -3 weeks lag (Train: 49w, Eval: 13w)}}} \\
\midrule
(i)  &            &             & $0.1262 \pm 2.78e-17$ \\
(ii) & \checkmark &             & $0.4024 \pm 5.55e-17$ \\
(iii)&            & \checkmark  & $0.1953 \pm 5.55e-17$ \\
(iv) & \checkmark & \checkmark  & $\mathbf{0.4176 \pm 5.55e-17}$ \\   
\bottomrule
\end{tabular}
\end{table}

Table~\ref{tab:ablation} presents the average value and the standard deviation of the models' $\mathrm{R}^2$ scores under these scenarios. In comparison to the baseline in (i), when either X posts count or Google Trends score are used as exogenous variables individually (ii-iii), we see a considerable increase in the average score. The biggest jump, however, is observed in (iv), where we employ both sources as exogenous variables, hinting at a complementary interaction between these two sources. All three latter scenarios show statistically significantly higher averages compared to the baseline scenario, according to a one-tailed Wilcoxon test, with $p$-values $< 0.001$.  {Additionally, we tested the effect of different lag structures for the exogenous variables by comparing models using unlagged data against those incorporating lags of $-1$, $-2$, and $-3$ weeks. Our results, presented in Table~\ref{tab:ablation}, show that incorporating a three-week lag for exogenous variables improved model performance, with the highest R² score observed in scenario (iv), where both X posts and Google Trends were included with a $-3$ week lag. Therefore, the results indicate that incorporating a three-week lag in exogenous variables enhances predictive accuracy compared to models using no lag or shorter lags.}

\section{Discussion}
\label{dis}

The primary objective of this study was to evaluate how effective the selected online data sources are in predicting AIV outbreaks. To assess the feasibility of this idea, we cross-correlated historical data from formal sources such as official notifications with informal ones such as X and Google Trends. This approach allowed us to capture temporal relationships between online activities and avian influenza reported cases to identify potential early indicators of AIV outbreaks. Additionally, a comparative analysis was conducted to identify reliable data sources between social media and Google Trends based on their performance across different locations.  {Moreover, an ablation study is performed to asses and isolate the positive effect of these informal sources in predicting AIV outbreaks.}

The collected historical data spans nearly two years and encompasses a wide set of languages. In this study, posts in French, Italian, Spanish, and Japanese were translated to English using an LLM before filtering out irrelevant posts using another model that was fine-tuned to 89.5\% accuracy. The utilized model in this study demonstrated a better performance in classifying relevant social media posts compared to the semi-supervised algorithm used in the previous study \citep{yousefinaghani2019assessment} that achieved an average accuracy score of 78.4\%. The superior performance of pre-trained models over traditional machine learning models could be attributed to their ability to leverage pre-existing knowledge and features from extensive prior training. Additionally, Canada and the USA had the largest number of social media posts and location data in our collected data. This is mostly due to the fact that the default language was English. However, the role of factors such as strict social media controls in certain countries, such as China, should not be underestimated. 

Social media post counts and search engine scores with officially reported case numbers were found highly correlated in several cases with high statistically significant coefficients, $1$-$2$ weeks prior to the start of the outbreak. Moreover, in regions with low levels of social media activity, search engine scores alone were still useful as an early indicator of an AIV outbreak, however, in those regions, confidence scores were lower. The SARIMAX ablation study further confirmed that combining X and Google Trends data provides the highest predictive accuracy, demonstrating the complementary strengths of the two sources. The lag analysis and the comparison of their varying correlation coefficients indicate that on a global level, Google Trends data was a better early indicator of avian influenza outbreaks. Moreover, findings from the selected locations in this study, specifically Canada and the USA, indicate that one province in Canada (BC) exhibited a statistically significant correlation. Yet, other provinces/states in both Canada and the USA still obtained relatively high correlation coefficients, but with $p$-values slightly higher than 5\%.

In Canada and the USA, where a closer inspection of outbreak waves is possible based on the focus of this study, Google Trends scores often peak before user activities on X and reported case numbers. This was the case for 3 (1st and 2nd Wave in Canada, and 2nd Wave in the USA) out of the 5 country-level waves under study. In the other two cases, X post counts reach their highest frequency first, followed closely by a rise in Google Trends scores and then reported case numbers. Another key observation is that for both Google Trends scores and social media activities, correlation coefficients were comparatively higher in country-level cases.

Variations in correlations and their statistical significance reflect the evolving dynamics of the outbreaks over time. There may be two reasons behind the diminishing correlations observed as the outbreaks progressed in subsequent weeks. Firstly, public interest or engagement in the outbreaks might have decreased which could result in reduced online activity as individuals felt less inclined to follow updates. Secondly, awareness and familiarity with avian influenza might have led to decreased online discussions or searches about the disease. Therefore, the approach proposed in this paper seems more suitable for detecting the initial outbreak waves and capturing the public's initial anxiety, curiosity, and interest in learning about the disease.

Despite the observed strengths in the utilized approach in this study compared to other existing works, several key challenges were encountered during the research process. These challenges were primarily concerned with data collection. First and foremost, restrictions that X recently imposed on their API endpoints complicated the process greatly. Moreover, the collected social media data was often riddled with inaccurate or missing location tags. Another notable limitation to social media data in some regions is due to limited access to these platforms or Internet access. Furthermore, while X does offer translations for non-English posts, they are not readily available via the API. In the present study, these challenges were addressed by employing LLMs to resolve missing and conflicting data points. Future research efforts may concentrate on improving the reliability of these data sources. However, despite the aforementioned challenges, the efficacy of these sources in real-time surveillance and their importance in developing early response strategies for avian influenza and other infectious diseases should not be underestimated.

\section{Conclusion}

\label{con}

In this study, an early identification mechanism using X and Google Trends was developed to identify avian influenza outbreak events, investigating the potential of online activities as supplementary sources of outbreak information. The results confirmed that this approach could complement traditional surveillance by providing timely information on avian influenza outbreaks. The findings presented here could also result in enhancing surveillance systems by offering early outbreak warnings and aiding animal health authorities in mitigating potential impacts. This study contributes to the literature by enhancing X post filtering for infectious disease and reducing noise using a large language model. Additionally, our ablation study further implies the predictive advantage of combining these data sources, demonstrating their complementary value in enhancing model performance. The principles applied here could benefit other animal infectious disease surveillance systems by reducing noise through post filtration.  {This methodology is, in fact, not limited to avian influenza outbreaks but could also be adapted to monitor other infectious diseases such as seasonal influenza, and emerging zoonotic diseases.} Future research could combine the two sources of data from social media and search engines or incorporate additional information such as repost and like counts for more precise results. However, challenges such as access to X's academic API and complexities in data translation and geolocation require technical and process related solutions to ensure accuracy. Enhancing data sources and collaboration among platforms, health authorities, and academia will be key to improving real-time outbreak surveillance and response strategies for avian influenza and other infectious diseases.

\subsection*{CRediT authorship contribution statement}

Marzieh Soltani: writing, original draft preparation, investigation, methodology, software, data curation, visualization,  validation, and interpretation of results. Rozita Dara: supervision, investigation, conceptualization, reviewing, editing, and validation. Zvonimir Poljak: reviewing and editing. Caroline Dubé: reviewing and editing. Neil Bruce: reviewing and editing. Shayan Sharif: supervision, conceptualization, reviewing, and editing.

\subsection*{Conflict of interest}

There is no conflict of interest in this study to declare.

\subsection*{Supplementary Material}

The Supplementary Material for this article can be found online at:

\subsection*{Acknowledgment}

This research is supported by the University of Guelph's Food from Thought initiative and Ontario Ministry of Agriculture Food and Rural Affairs, Alliance Tier I, funding.

\subsection*{Data availability}
Data will be made available on request.

\printcredits


\begin{thebibliography}{}

\bibitem [\protect \citeauthoryear {%
AbuBakar%
\ \protect \BOthers {.}}{%
AbuBakar%
\ \protect \BOthers {.}}{%
{\protect \APACyear {2023}}%
}]{%
abubakar2023avian}
\APACinsertmetastar {%
abubakar2023avian}%
\begin{APACrefauthors}%
AbuBakar, U.%
, Amrani, L.%
, Kamarulzaman, F\BPBI A.%
, Karsani, S\BPBI A.%
, Hassandarvish, P.%
\BCBL {}\ \BBA {} Khairat, J\BPBI E.%
\end{APACrefauthors}%
\unskip\
\newblock
\APACrefYearMonthDay{2023}{}{}.
\newblock
{\BBOQ}\APACrefatitle {Avian influenza virus tropism in humans} {Avian influenza virus tropism in humans}.{\BBCQ}
\newblock
\APACjournalVolNumPages{Viruses}{15}{4}{833}.
\PrintBackRefs{\CurrentBib}

\bibitem [\protect \citeauthoryear {%
Adams%
\ \protect \BOthers {.}}{%
Adams%
\ \protect \BOthers {.}}{%
{\protect \APACyear {2022}}%
}]{%
adams2022}
\APACinsertmetastar {%
adams2022}%
\begin{APACrefauthors}%
Adams, C.%
, Bozhidarova, M.%
, Chen, J.%
, Gao, A.%
, Liu, Z.%
, Priniski, J\BPBI H.%
\BDBL {}Brantingham, P\BPBI J.%
\end{APACrefauthors}%
\unskip\
\newblock
\APACrefYearMonthDay{2022}{}{}.
\newblock
{\BBOQ}\APACrefatitle {Knowledge graphs of the QAnon Twitter network} {Knowledge graphs of the qanon twitter network}.{\BBCQ}
\newblock
\BIn{} \APACrefbtitle {2022 IEEE International Conference on Big Data (Big Data)} {2022 ieee international conference on big data (big data)}\ (\BPGS\ 2903--2912).
\PrintBackRefs{\CurrentBib}

\bibitem [\protect \citeauthoryear {%
Ahmed%
, Bath%
, Sbaffi%
\BCBL {}\ \BBA {} Demartini%
}{%
Ahmed%
\ \protect \BOthers {.}}{%
{\protect \APACyear {2018}}%
}]{%
ahmed2018using}
\APACinsertmetastar {%
ahmed2018using}%
\begin{APACrefauthors}%
Ahmed, W.%
, Bath, P.%
, Sbaffi, L.%
\BCBL {}\ \BBA {} Demartini, G.%
\end{APACrefauthors}%
\unskip\
\newblock
\APACrefYearMonthDay{2018}{}{}.
\newblock
{\BBOQ}\APACrefatitle {Using Twitter for insights into the 2009 swine flu and 2014 Ebola outbreaks} {Using twitter for insights into the 2009 swine flu and 2014 ebola outbreaks}.{\BBCQ}
\newblock

\PrintBackRefs{\CurrentBib}

\bibitem [\protect \citeauthoryear {%
Alkouz%
, Al~Aghbari%
, Al-Garadi%
\BCBL {}\ \BBA {} Sarker%
}{%
Alkouz%
\ \protect \BOthers {.}}{%
{\protect \APACyear {2022}}%
}]{%
alkouz2022deepluenza}
\APACinsertmetastar {%
alkouz2022deepluenza}%
\begin{APACrefauthors}%
Alkouz, B.%
, Al~Aghbari, Z.%
, Al-Garadi, M\BPBI A.%
\BCBL {}\ \BBA {} Sarker, A.%
\end{APACrefauthors}%
\unskip\
\newblock
\APACrefYearMonthDay{2022}{}{}.
\newblock
{\BBOQ}\APACrefatitle {Deepluenza: Deep learning for influenza detection from twitter} {Deepluenza: Deep learning for influenza detection from twitter}.{\BBCQ}
\newblock
\APACjournalVolNumPages{Expert Systems with Applications}{198}{}{116845}.
\PrintBackRefs{\CurrentBib}

\bibitem [\protect \citeauthoryear {%
Arinik%
, Interdonato%
, Roche%
\BCBL {}\ \BBA {} Teisseire%
}{%
Arinik%
\ \protect \BOthers {.}}{%
{\protect \APACyear {2023}}%
}]{%
arinik2023evaluation}
\APACinsertmetastar {%
arinik2023evaluation}%
\begin{APACrefauthors}%
Arinik, N.%
, Interdonato, R.%
, Roche, M.%
\BCBL {}\ \BBA {} Teisseire, M.%
\end{APACrefauthors}%
\unskip\
\newblock
\APACrefYearMonthDay{2023}{}{}.
\newblock
{\BBOQ}\APACrefatitle {An evaluation framework for comparing epidemic intelligence systems} {An evaluation framework for comparing epidemic intelligence systems}.{\BBCQ}
\newblock
\APACjournalVolNumPages{IEEE Access}{11}{}{31880--31901}.
\PrintBackRefs{\CurrentBib}

\bibitem [\protect \citeauthoryear {%
Bahk%
, Scales%
, Mekaru%
, Brownstein%
\BCBL {}\ \BBA {} Freifeld%
}{%
Bahk%
\ \protect \BOthers {.}}{%
{\protect \APACyear {2015}}%
}]{%
bahk2015comparing}
\APACinsertmetastar {%
bahk2015comparing}%
\begin{APACrefauthors}%
Bahk, C\BPBI Y.%
, Scales, D\BPBI A.%
, Mekaru, S\BPBI R.%
, Brownstein, J\BPBI S.%
\BCBL {}\ \BBA {} Freifeld, C\BPBI C.%
\end{APACrefauthors}%
\unskip\
\newblock
\APACrefYearMonthDay{2015}{}{}.
\newblock
{\BBOQ}\APACrefatitle {Comparing timeliness, content, and disease severity of formal and informal source outbreak reporting} {Comparing timeliness, content, and disease severity of formal and informal source outbreak reporting}.{\BBCQ}
\newblock
\APACjournalVolNumPages{BMC infectious diseases}{15}{1}{1--6}.
\PrintBackRefs{\CurrentBib}

\bibitem [\protect \citeauthoryear {%
Barboza%
\ \protect \BOthers {.}}{%
Barboza%
\ \protect \BOthers {.}}{%
{\protect \APACyear {2013}}%
}]{%
barboza2013evaluation}
\APACinsertmetastar {%
barboza2013evaluation}%
\begin{APACrefauthors}%
Barboza, P.%
, Vaillant, L.%
, Mawudeku, A.%
, Nelson, N\BPBI P.%
, Hartley, D\BPBI M.%
, Madoff, L\BPBI C.%
\BDBL {}others%
\end{APACrefauthors}%
\unskip\
\newblock
\APACrefYearMonthDay{2013}{}{}.
\newblock
{\BBOQ}\APACrefatitle {Evaluation of epidemic intelligence systems integrated in the early alerting and reporting project for the detection of A/H5N1 influenza events} {Evaluation of epidemic intelligence systems integrated in the early alerting and reporting project for the detection of a/h5n1 influenza events}.{\BBCQ}
\newblock
\APACjournalVolNumPages{PLoS One}{8}{3}{e57252}.
\PrintBackRefs{\CurrentBib}

\bibitem [\protect \citeauthoryear {%
Bernardo%
\ \protect \BOthers {.}}{%
Bernardo%
\ \protect \BOthers {.}}{%
{\protect \APACyear {2013}}%
}]{%
bernardo2013scoping}
\APACinsertmetastar {%
bernardo2013scoping}%
\begin{APACrefauthors}%
Bernardo, T\BPBI M.%
, Rajic, A.%
, Young, I.%
, Robiadek, K.%
, Pham, M\BPBI T.%
\BCBL {}\ \BBA {} Funk, J\BPBI A.%
\end{APACrefauthors}%
\unskip\
\newblock
\APACrefYearMonthDay{2013}{}{}.
\newblock
{\BBOQ}\APACrefatitle {Scoping review on search queries and social media for disease surveillance: a chronology of innovation} {Scoping review on search queries and social media for disease surveillance: a chronology of innovation}.{\BBCQ}
\newblock
\APACjournalVolNumPages{Journal of medical Internet research}{15}{7}{e2740}.
\PrintBackRefs{\CurrentBib}

\bibitem [\protect \citeauthoryear {%
Blagodatski%
\ \protect \BOthers {.}}{%
Blagodatski%
\ \protect \BOthers {.}}{%
{\protect \APACyear {2021}}%
}]{%
blagodatski2021avian}
\APACinsertmetastar {%
blagodatski2021avian}%
\begin{APACrefauthors}%
Blagodatski, A.%
, Trutneva, K.%
, Glazova, O.%
, Mityaeva, O.%
, Shevkova, L.%
, Kegeles, E.%
\BDBL {}others%
\end{APACrefauthors}%
\unskip\
\newblock
\APACrefYearMonthDay{2021}{}{}.
\newblock
{\BBOQ}\APACrefatitle {Avian influenza in wild birds and poultry: dissemination pathways, monitoring methods, and virus ecology} {Avian influenza in wild birds and poultry: dissemination pathways, monitoring methods, and virus ecology}.{\BBCQ}
\newblock
\APACjournalVolNumPages{Pathogens}{10}{5}{630}.
\PrintBackRefs{\CurrentBib}

\bibitem [\protect \citeauthoryear {%
Bommasani%
\ \protect \BOthers {.}}{%
Bommasani%
\ \protect \BOthers {.}}{%
{\protect \APACyear {2021}}%
}]{%
bommasani2021opportunities}
\APACinsertmetastar {%
bommasani2021opportunities}%
\begin{APACrefauthors}%
Bommasani, R.%
, Hudson, D\BPBI A.%
, Adeli, E.%
, Altman, R.%
, Arora, S.%
, von Arx, S.%
\BDBL {}others%
\end{APACrefauthors}%
\unskip\
\newblock
\APACrefYearMonthDay{2021}{}{}.
\newblock
{\BBOQ}\APACrefatitle {On the opportunities and risks of foundation models} {On the opportunities and risks of foundation models}.{\BBCQ}
\newblock
\APACjournalVolNumPages{arXiv preprint arXiv:2108.07258}{}{}{}.
\PrintBackRefs{\CurrentBib}

\bibitem [\protect \citeauthoryear {%
Box%
, Jenkins%
, Reinsel%
\BCBL {}\ \BBA {} Ljung%
}{%
Box%
\ \protect \BOthers {.}}{%
{\protect \APACyear {2015}}%
}]{%
box2015time}
\APACinsertmetastar {%
box2015time}%
\begin{APACrefauthors}%
Box, G\BPBI E.%
, Jenkins, G\BPBI M.%
, Reinsel, G\BPBI C.%
\BCBL {}\ \BBA {} Ljung, G\BPBI M.%
\end{APACrefauthors}%
\unskip\
\newblock
\APACrefYear{2015}.
\newblock
\APACrefbtitle {Time series analysis: forecasting and control} {Time series analysis: forecasting and control}.
\newblock
\APACaddressPublisher{}{John Wiley \& Sons}.
\PrintBackRefs{\CurrentBib}

\bibitem [\protect \citeauthoryear {%
{CCDB}%
}{%
{CCDB}%
}{%
{\protect \APACyear {{\protect \bibnodate {}}}}%
}]{%
Digital}
\APACinsertmetastar {%
Digital}%
\begin{APACrefauthors}%
{CCDB}.%
\end{APACrefauthors}%
\unskip\
\newblock
\APACrefYearMonthDay{{\protect \bibnodate {}}}{}{}.
\newblock
\APACrefbtitle {Digital Research Alliance of Canada.} {Digital research alliance of canada.}
\newblock
\APAChowpublished {\url{https://alliancecan.ca/en//}}.
\newblock
\APACrefnote{Accessed: 2024}
\PrintBackRefs{\CurrentBib}

\bibitem [\protect \citeauthoryear {%
Chen%
\ \protect \BOthers {.}}{%
Chen%
\ \protect \BOthers {.}}{%
{\protect \APACyear {2019}}%
}]{%
chen2019avian}
\APACinsertmetastar {%
chen2019avian}%
\begin{APACrefauthors}%
Chen, Y.%
, Zhang, Y.%
, Xu, Z.%
, Wang, X.%
, Lu, J.%
\BCBL {}\ \BBA {} Hu, W.%
\end{APACrefauthors}%
\unskip\
\newblock
\APACrefYearMonthDay{2019}{}{}.
\newblock
{\BBOQ}\APACrefatitle {Avian Influenza A (H7N9) and related Internet search query data in China} {Avian influenza a (h7n9) and related internet search query data in china}.{\BBCQ}
\newblock
\APACjournalVolNumPages{Scientific reports}{9}{1}{10434}.
\PrintBackRefs{\CurrentBib}

\bibitem [\protect \citeauthoryear {%
Christaki%
}{%
Christaki%
}{%
{\protect \APACyear {2015}}%
}]{%
christaki2015new}
\APACinsertmetastar {%
christaki2015new}%
\begin{APACrefauthors}%
Christaki, E.%
\end{APACrefauthors}%
\unskip\
\newblock
\APACrefYearMonthDay{2015}{}{}.
\newblock
{\BBOQ}\APACrefatitle {New technologies in predicting, preventing and controlling emerging infectious diseases} {New technologies in predicting, preventing and controlling emerging infectious diseases}.{\BBCQ}
\newblock
\APACjournalVolNumPages{Virulence}{6}{6}{558--565}.
\PrintBackRefs{\CurrentBib}

\bibitem [\protect \citeauthoryear {%
Costa-juss{\`a}%
\ \protect \BOthers {.}}{%
Costa-juss{\`a}%
\ \protect \BOthers {.}}{%
{\protect \APACyear {2022}}%
}]{%
team2022}
\APACinsertmetastar {%
team2022}%
\begin{APACrefauthors}%
Costa-juss{\`a}, M\BPBI R.%
, Cross, J.%
, {\c{C}}elebi, O.%
, Elbayad, M.%
, Heafield, K.%
, Heffernan, K.%
\BDBL {}others%
\end{APACrefauthors}%
\unskip\
\newblock
\APACrefYearMonthDay{2022}{}{}.
\newblock
{\BBOQ}\APACrefatitle {No language left behind: Scaling human-centered machine translation} {No language left behind: Scaling human-centered machine translation}.{\BBCQ}
\newblock
\APACjournalVolNumPages{arXiv preprint arXiv:2207.04672}{}{}{}.
\PrintBackRefs{\CurrentBib}

\bibitem [\protect \citeauthoryear {%
Culotta%
}{%
Culotta%
}{%
{\protect \APACyear {2010}}%
}]{%
culotta2010towards}
\APACinsertmetastar {%
culotta2010towards}%
\begin{APACrefauthors}%
Culotta, A.%
\end{APACrefauthors}%
\unskip\
\newblock
\APACrefYearMonthDay{2010}{}{}.
\newblock
{\BBOQ}\APACrefatitle {Towards detecting influenza epidemics by analyzing Twitter messages} {Towards detecting influenza epidemics by analyzing twitter messages}.{\BBCQ}
\newblock
\BIn{} \APACrefbtitle {Proceedings of the first workshop on social media analytics} {Proceedings of the first workshop on social media analytics}\ (\BPGS\ 115--122).
\PrintBackRefs{\CurrentBib}

\bibitem [\protect \citeauthoryear {%
Deiner%
\ \protect \BOthers {.}}{%
Deiner%
\ \protect \BOthers {.}}{%
{\protect \APACyear {2024}}%
}]{%
deiner2024use}
\APACinsertmetastar {%
deiner2024use}%
\begin{APACrefauthors}%
Deiner, M\BPBI S.%
, Deiner, N\BPBI A.%
, Hristidis, V.%
, McLeod, S\BPBI D.%
, Doan, T.%
, Lietman, T\BPBI M.%
\BCBL {}\ \BBA {} Porco, T\BPBI C.%
\end{APACrefauthors}%
\unskip\
\newblock
\APACrefYearMonthDay{2024}{}{}.
\newblock
{\BBOQ}\APACrefatitle {Use of Large Language Models to Assess the Likelihood of Epidemics From the Content of Tweets: Infodemiology Study} {Use of large language models to assess the likelihood of epidemics from the content of tweets: Infodemiology study}.{\BBCQ}
\newblock
\APACjournalVolNumPages{Journal of Medical Internet Research}{26}{}{e49139}.
\PrintBackRefs{\CurrentBib}

\bibitem [\protect \citeauthoryear {%
Devlin%
, Chang%
, Lee%
\BCBL {}\ \BBA {} Toutanova%
}{%
Devlin%
\ \protect \BOthers {.}}{%
{\protect \APACyear {2018}}%
}]{%
devlin2018bert}
\APACinsertmetastar {%
devlin2018bert}%
\begin{APACrefauthors}%
Devlin, J.%
, Chang, M\BHBI W.%
, Lee, K.%
\BCBL {}\ \BBA {} Toutanova, K.%
\end{APACrefauthors}%
\unskip\
\newblock
\APACrefYearMonthDay{2018}{}{}.
\newblock
{\BBOQ}\APACrefatitle {Bert: Pre-training of deep bidirectional transformers for language understanding} {Bert: Pre-training of deep bidirectional transformers for language understanding}.{\BBCQ}
\newblock
\APACjournalVolNumPages{arXiv preprint arXiv:1810.04805}{}{}{}.
\PrintBackRefs{\CurrentBib}

\bibitem [\protect \citeauthoryear {%
Di~Martino%
\ \protect \BOthers {.}}{%
Di~Martino%
\ \protect \BOthers {.}}{%
{\protect \APACyear {2017}}%
}]{%
dimartino2017}
\APACinsertmetastar {%
dimartino2017}%
\begin{APACrefauthors}%
Di~Martino, S.%
, Romano, S.%
, Bertolotto, M.%
, Kanhabua, N.%
, Mazzeo, A.%
\BCBL {}\ \BBA {} Nejdl, W.%
\end{APACrefauthors}%
\unskip\
\newblock
\APACrefYearMonthDay{2017}{}{}.
\newblock
{\BBOQ}\APACrefatitle {Towards exploiting social networks for detecting epidemic outbreaks} {Towards exploiting social networks for detecting epidemic outbreaks}.{\BBCQ}
\newblock
\APACjournalVolNumPages{Global Journal of Flexible Systems Management}{18}{}{61--71}.
\PrintBackRefs{\CurrentBib}

\bibitem [\protect \citeauthoryear {%
Dion%
, AbdelMalik%
\BCBL {}\ \BBA {} Mawudeku%
}{%
Dion%
\ \protect \BOthers {.}}{%
{\protect \APACyear {2015}}%
}]{%
dion2015big}
\APACinsertmetastar {%
dion2015big}%
\begin{APACrefauthors}%
Dion, M.%
, AbdelMalik, P.%
\BCBL {}\ \BBA {} Mawudeku, A.%
\end{APACrefauthors}%
\unskip\
\newblock
\APACrefYearMonthDay{2015}{}{}.
\newblock
{\BBOQ}\APACrefatitle {Big data: big data and the global public health intelligence network (GPHIN)} {Big data: big data and the global public health intelligence network (gphin)}.{\BBCQ}
\newblock
\APACjournalVolNumPages{Canada Communicable Disease Report}{41}{9}{209}.
\PrintBackRefs{\CurrentBib}

\bibitem [\protect \citeauthoryear {%
Duan%
, Li%
, Ren%
, Bai%
\BCBL {}\ \BBA {} Zhou%
}{%
Duan%
\ \protect \BOthers {.}}{%
{\protect \APACyear {2023}}%
}]{%
duan2023overview}
\APACinsertmetastar {%
duan2023overview}%
\begin{APACrefauthors}%
Duan, C.%
, Li, C.%
, Ren, R.%
, Bai, W.%
\BCBL {}\ \BBA {} Zhou, L.%
\end{APACrefauthors}%
\unskip\
\newblock
\APACrefYearMonthDay{2023}{}{}.
\newblock
{\BBOQ}\APACrefatitle {An overview of avian influenza surveillance strategies and modes} {An overview of avian influenza surveillance strategies and modes}.{\BBCQ}
\newblock
\APACjournalVolNumPages{Science in One Health}{}{}{100043}.
\PrintBackRefs{\CurrentBib}

\bibitem [\protect \citeauthoryear {%
Durbin%
\ \BBA {} Koopman%
}{%
Durbin%
\ \BBA {} Koopman%
}{%
{\protect \APACyear {2012}}%
}]{%
durbin2012time}
\APACinsertmetastar {%
durbin2012time}%
\begin{APACrefauthors}%
Durbin, J.%
\BCBT {}\ \BBA {} Koopman, S\BPBI J.%
\end{APACrefauthors}%
\unskip\
\newblock
\APACrefYear{2012}.
\newblock
\APACrefbtitle {Time series analysis by state space methods} {Time series analysis by state space methods}\ (\BVOL~38).
\newblock
\APACaddressPublisher{}{OUP Oxford}.
\PrintBackRefs{\CurrentBib}

\bibitem [\protect \citeauthoryear {%
Fast%
\ \protect \BOthers {.}}{%
Fast%
\ \protect \BOthers {.}}{%
{\protect \APACyear {2018}}%
}]{%
fast2018predicting}
\APACinsertmetastar {%
fast2018predicting}%
\begin{APACrefauthors}%
Fast, S\BPBI M.%
, Kim, L.%
, Cohn, E\BPBI L.%
, Mekaru, S\BPBI R.%
, Brownstein, J\BPBI S.%
\BCBL {}\ \BBA {} Markuzon, N.%
\end{APACrefauthors}%
\unskip\
\newblock
\APACrefYearMonthDay{2018}{}{}.
\newblock
{\BBOQ}\APACrefatitle {Predicting social response to infectious disease outbreaks from internet-based news streams} {Predicting social response to infectious disease outbreaks from internet-based news streams}.{\BBCQ}
\newblock
\APACjournalVolNumPages{Annals of Operations Research}{263}{}{551--564}.
\PrintBackRefs{\CurrentBib}

\bibitem [\protect \citeauthoryear {%
Ginsberg%
\ \protect \BOthers {.}}{%
Ginsberg%
\ \protect \BOthers {.}}{%
{\protect \APACyear {2009}}%
}]{%
ginsberg2009detecting}
\APACinsertmetastar {%
ginsberg2009detecting}%
\begin{APACrefauthors}%
Ginsberg, J.%
, Mohebbi, M\BPBI H.%
, Patel, R\BPBI S.%
, Brammer, L.%
, Smolinski, M\BPBI S.%
\BCBL {}\ \BBA {} Brilliant, L.%
\end{APACrefauthors}%
\unskip\
\newblock
\APACrefYearMonthDay{2009}{}{}.
\newblock
{\BBOQ}\APACrefatitle {Detecting influenza epidemics using search engine query data} {Detecting influenza epidemics using search engine query data}.{\BBCQ}
\newblock
\APACjournalVolNumPages{Nature}{457}{7232}{1012--1014}.
\PrintBackRefs{\CurrentBib}

\bibitem [\protect \citeauthoryear {%
Google%
}{%
Google%
}{%
{\protect \APACyear {Accessed 2023}}%
}]{%
GoogleTrends}
\APACinsertmetastar {%
GoogleTrends}%
\begin{APACrefauthors}%
Google.%
\end{APACrefauthors}%
\unskip\
\newblock
\APACrefYearMonthDay{Accessed 2023}{}{}.
\newblock
\APACrefbtitle {Google Trends Help Center.} {Google trends help center.}
\newblock
\begin{APACrefURL} \url{https://support.google.com/trends/answer/4365533?hl=en&ref_topic=6248052&sjid=4115413422933437318-NA} \end{APACrefURL}
\PrintBackRefs{\CurrentBib}

\bibitem [\protect \citeauthoryear {%
Graham%
\ \protect \BOthers {.}}{%
Graham%
\ \protect \BOthers {.}}{%
{\protect \APACyear {2018}}%
}]{%
graham2018prepared}
\APACinsertmetastar {%
graham2018prepared}%
\begin{APACrefauthors}%
Graham, J\BPBI E.%
, Lees, S.%
, Le~Marcis, F.%
, Faye, S\BPBI L.%
, Lorway, R\BPBI R.%
, Ronse, M.%
\BDBL {}Grietens, K\BPBI P.%
\end{APACrefauthors}%
\unskip\
\newblock
\APACrefYearMonthDay{2018}{}{}.
\newblock
{\BBOQ}\APACrefatitle {Prepared for the ‘unexpected’? Lessons from the 2014--2016 Ebola epidemic in West Africa on integrating emergent theory designs into outbreak response} {Prepared for the ‘unexpected’? lessons from the 2014--2016 ebola epidemic in west africa on integrating emergent theory designs into outbreak response}.{\BBCQ}
\newblock
\APACjournalVolNumPages{BMJ global health}{3}{4}{}.
\PrintBackRefs{\CurrentBib}

\bibitem [\protect \citeauthoryear {%
Han%
\ \protect \BOthers {.}}{%
Han%
\ \protect \BOthers {.}}{%
{\protect \APACyear {2021}}%
}]{%
han2021pre}
\APACinsertmetastar {%
han2021pre}%
\begin{APACrefauthors}%
Han, X.%
, Zhang, Z.%
, Ding, N.%
, Gu, Y.%
, Liu, X.%
, Huo, Y.%
\BDBL {}others%
\end{APACrefauthors}%
\unskip\
\newblock
\APACrefYearMonthDay{2021}{}{}.
\newblock
{\BBOQ}\APACrefatitle {Pre-trained models: Past, present and future} {Pre-trained models: Past, present and future}.{\BBCQ}
\newblock
\APACjournalVolNumPages{AI Open}{2}{}{225--250}.
\PrintBackRefs{\CurrentBib}

\bibitem [\protect \citeauthoryear {%
{Hugging Face}%
}{%
{Hugging Face}%
}{%
{\protect \APACyear {2023}}%
}]{%
huggingface}
\APACinsertmetastar {%
huggingface}%
\begin{APACrefauthors}%
{Hugging Face}.%
\end{APACrefauthors}%
\unskip\
\newblock
\APACrefYearMonthDay{2023}{}{}.
\newblock
\APACrefbtitle {facebook/nllb-200-3.3B: A Large Multilingual Transformer Language Model.} {facebook/nllb-200-3.3b: A large multilingual transformer language model.}
\newblock
\APAChowpublished {\url{https://huggingface.co/facebook/nllb-200-3.3B}}.
\newblock
\APACrefnote{Accessed: Date}
\PrintBackRefs{\CurrentBib}

\bibitem [\protect \citeauthoryear {%
Johnson%
, Seeger%
\BCBL {}\ \BBA {} Marsh%
}{%
Johnson%
\ \protect \BOthers {.}}{%
{\protect \APACyear {2016}}%
}]{%
johnson2016local}
\APACinsertmetastar {%
johnson2016local}%
\begin{APACrefauthors}%
Johnson, K\BPBI K.%
, Seeger, R\BPBI M.%
\BCBL {}\ \BBA {} Marsh, T\BPBI L.%
\end{APACrefauthors}%
\unskip\
\newblock
\APACrefYearMonthDay{2016}{}{}.
\newblock
{\BBOQ}\APACrefatitle {Local economies and highly pathogenic avian influenza} {Local economies and highly pathogenic avian influenza}.{\BBCQ}
\newblock
\APACjournalVolNumPages{Choices}{31}{2}{1--9}.
\PrintBackRefs{\CurrentBib}

\bibitem [\protect \citeauthoryear {%
Kaiser%
, Coulombier%
, Baldari%
, Morgan%
\BCBL {}\ \BBA {} Paquet%
}{%
Kaiser%
\ \protect \BOthers {.}}{%
{\protect \APACyear {2006}}%
}]{%
kaiser2006epidemic}
\APACinsertmetastar {%
kaiser2006epidemic}%
\begin{APACrefauthors}%
Kaiser, R.%
, Coulombier, D.%
, Baldari, M.%
, Morgan, D.%
\BCBL {}\ \BBA {} Paquet, C.%
\end{APACrefauthors}%
\unskip\
\newblock
\APACrefYearMonthDay{2006}{}{}.
\newblock
{\BBOQ}\APACrefatitle {What is epidemic intelligence, and how is it being improved in Europe?} {What is epidemic intelligence, and how is it being improved in europe?}{\BBCQ}
\newblock
\APACjournalVolNumPages{Weekly releases (1997--2007)}{11}{5}{2892}.
\PrintBackRefs{\CurrentBib}

\bibitem [\protect \citeauthoryear {%
Kapit{\'a}ny-F{\"o}v{\'e}ny%
\ \protect \BOthers {.}}{%
Kapit{\'a}ny-F{\"o}v{\'e}ny%
\ \protect \BOthers {.}}{%
{\protect \APACyear {2019}}%
}]{%
kapitany2019can}
\APACinsertmetastar {%
kapitany2019can}%
\begin{APACrefauthors}%
Kapit{\'a}ny-F{\"o}v{\'e}ny, M.%
, Ferenci, T.%
, Sulyok, Z.%
, Kegele, J.%
, Richter, H.%
, V{\'a}lyi-Nagy, I.%
\BCBL {}\ \BBA {} Sulyok, M.%
\end{APACrefauthors}%
\unskip\
\newblock
\APACrefYearMonthDay{2019}{}{}.
\newblock
{\BBOQ}\APACrefatitle {Can Google Trends data improve forecasting of Lyme disease incidence?} {Can google trends data improve forecasting of lyme disease incidence?}{\BBCQ}
\newblock
\APACjournalVolNumPages{Zoonoses and public health}{66}{1}{101--107}.
\PrintBackRefs{\CurrentBib}

\bibitem [\protect \citeauthoryear {%
Leguia%
\ \protect \BOthers {.}}{%
Leguia%
\ \protect \BOthers {.}}{%
{\protect \APACyear {2023}}%
}]{%
leguia2023highly}
\APACinsertmetastar {%
leguia2023highly}%
\begin{APACrefauthors}%
Leguia, M.%
, Garcia-Glaessner, A.%
, Mu{\~n}oz-Saavedra, B.%
, Juarez, D.%
, Barrera, P.%
, Calvo-Mac, C.%
\BDBL {}others%
\end{APACrefauthors}%
\unskip\
\newblock
\APACrefYearMonthDay{2023}{}{}.
\newblock
{\BBOQ}\APACrefatitle {Highly pathogenic avian influenza A (H5N1) in marine mammals and seabirds in Peru} {Highly pathogenic avian influenza a (h5n1) in marine mammals and seabirds in peru}.{\BBCQ}
\newblock
\APACjournalVolNumPages{Nature Communications}{14}{1}{5489}.
\PrintBackRefs{\CurrentBib}

\bibitem [\protect \citeauthoryear {%
Liu%
, Feng%
, Tsui%
\BCBL {}\ \BBA {} Sun%
}{%
Liu%
\ \protect \BOthers {.}}{%
{\protect \APACyear {2021}}%
}]{%
liu2021forecasting}
\APACinsertmetastar {%
liu2021forecasting}%
\begin{APACrefauthors}%
Liu, Y.%
, Feng, G.%
, Tsui, K\BHBI L.%
\BCBL {}\ \BBA {} Sun, S.%
\end{APACrefauthors}%
\unskip\
\newblock
\APACrefYearMonthDay{2021}{}{}.
\newblock
{\BBOQ}\APACrefatitle {Forecasting influenza epidemics in Hong Kong using Google search queries data: A new integrated approach} {Forecasting influenza epidemics in hong kong using google search queries data: A new integrated approach}.{\BBCQ}
\newblock
\APACjournalVolNumPages{Expert Systems with Applications}{185}{}{115604}.
\PrintBackRefs{\CurrentBib}

\bibitem [\protect \citeauthoryear {%
Lu%
\ \protect \BOthers {.}}{%
Lu%
\ \protect \BOthers {.}}{%
{\protect \APACyear {2019}}%
}]{%
lu2018}
\APACinsertmetastar {%
lu2018}%
\begin{APACrefauthors}%
Lu, Y.%
, Wang, S.%
, Wang, J.%
, Zhou, G.%
, Zhang, Q.%
, Zhou, X.%
\BDBL {}Chou, K\BHBI C.%
\end{APACrefauthors}%
\unskip\
\newblock
\APACrefYearMonthDay{2019}{{\APACmonth{03}}}{}.
\newblock
{\BBOQ}\APACrefatitle {An Epidemic Avian Influenza Prediction Model Based on Google Trends} {An epidemic avian influenza prediction model based on google trends}.{\BBCQ}
\newblock
\APACjournalVolNumPages{Letters in Organic Chemistry}{16}{4}{303–310}.
\newblock
\begin{APACrefURL} \url{http://dx.doi.org/10.2174/1570178615666180724103325} \end{APACrefURL}
\newblock
\begin{APACrefDOI} \doi{10.2174/1570178615666180724103325} \end{APACrefDOI}
\PrintBackRefs{\CurrentBib}

\bibitem [\protect \citeauthoryear {%
Lyon%
}{%
Lyon%
}{%
{\protect \APACyear {2010}}%
}]{%
lyon2010discrete}
\APACinsertmetastar {%
lyon2010discrete}%
\begin{APACrefauthors}%
Lyon, D.%
\end{APACrefauthors}%
\unskip\
\newblock
\APACrefYearMonthDay{2010}{}{}.
\newblock
{\BBOQ}\APACrefatitle {The Discrete Fourier Transform, part 6: Cross-correlation.} {The discrete fourier transform, part 6: Cross-correlation.}{\BBCQ}
\newblock
\APACjournalVolNumPages{J. Object Technol.}{9}{2}{17--22}.
\PrintBackRefs{\CurrentBib}

\bibitem [\protect \citeauthoryear {%
Moorhead%
\ \protect \BOthers {.}}{%
Moorhead%
\ \protect \BOthers {.}}{%
{\protect \APACyear {2013}}%
}]{%
moorhead2013new}
\APACinsertmetastar {%
moorhead2013new}%
\begin{APACrefauthors}%
Moorhead, S\BPBI A.%
, Hazlett, D\BPBI E.%
, Harrison, L.%
, Carroll, J\BPBI K.%
, Irwin, A.%
\BCBL {}\ \BBA {} Hoving, C.%
\end{APACrefauthors}%
\unskip\
\newblock
\APACrefYearMonthDay{2013}{}{}.
\newblock
{\BBOQ}\APACrefatitle {A new dimension of health care: systematic review of the uses, benefits, and limitations of social media for health communication} {A new dimension of health care: systematic review of the uses, benefits, and limitations of social media for health communication}.{\BBCQ}
\newblock
\APACjournalVolNumPages{Journal of medical Internet research}{15}{4}{e1933}.
\PrintBackRefs{\CurrentBib}

\bibitem [\protect \citeauthoryear {%
Morsy%
\ \protect \BOthers {.}}{%
Morsy%
\ \protect \BOthers {.}}{%
{\protect \APACyear {2018}}%
}]{%
morsy2018prediction}
\APACinsertmetastar {%
morsy2018prediction}%
\begin{APACrefauthors}%
Morsy, S.%
, Dang, T.%
, Kamel, M.%
, Zayan, A.%
, Makram, O.%
, Elhady, M.%
\BDBL {}Huy, N.%
\end{APACrefauthors}%
\unskip\
\newblock
\APACrefYearMonthDay{2018}{}{}.
\newblock
{\BBOQ}\APACrefatitle {Prediction of Zika-confirmed cases in Brazil and Colombia using Google Trends} {Prediction of zika-confirmed cases in brazil and colombia using google trends}.{\BBCQ}
\newblock
\APACjournalVolNumPages{Epidemiology \& Infection}{146}{13}{1625--1627}.
\PrintBackRefs{\CurrentBib}

\bibitem [\protect \citeauthoryear {%
Mounica%
\ \BBA {} Lavanya%
}{%
Mounica%
\ \BBA {} Lavanya%
}{%
{\protect \APACyear {2024}}%
}]{%
mounica2022}
\APACinsertmetastar {%
mounica2022}%
\begin{APACrefauthors}%
Mounica, B.%
\BCBT {}\ \BBA {} Lavanya, K.%
\end{APACrefauthors}%
\unskip\
\newblock
\APACrefYearMonthDay{2024}{}{}.
\newblock
{\BBOQ}\APACrefatitle {Feature selection method on twitter dataset with part-of-speech (PoS) pattern applied to traffic analysis} {Feature selection method on twitter dataset with part-of-speech (pos) pattern applied to traffic analysis}.{\BBCQ}
\newblock
\APACjournalVolNumPages{International Journal of System Assurance Engineering and Management}{15}{1}{110--123}.
\PrintBackRefs{\CurrentBib}

\bibitem [\protect \citeauthoryear {%
Organization%
\ \protect \BOthers {.}}{%
Organization%
\ \protect \BOthers {.}}{%
{\protect \APACyear {{\protect \bibnodate {}}}}%
}]{%
worldongoing}
\APACinsertmetastar {%
worldongoing}%
\begin{APACrefauthors}%
Organization, W\BPBI H.%
\BCBT {}\ \BOthersPeriod {.}
\end{APACrefauthors}%
\unskip\
\newblock
\APACrefYearMonthDay{{\protect \bibnodate {}}}{}{}.
\newblock
\APACrefbtitle {Ongoing avian influenza outbreaks in animals pose risk to humans [cited 2023 July 12].} {Ongoing avian influenza outbreaks in animals pose risk to humans [cited 2023 july 12].}
\PrintBackRefs{\CurrentBib}

\bibitem [\protect \citeauthoryear {%
Pandya%
\ \BBA {} Lodha%
}{%
Pandya%
\ \BBA {} Lodha%
}{%
{\protect \APACyear {2021}}%
}]{%
pandya2021social}
\APACinsertmetastar {%
pandya2021social}%
\begin{APACrefauthors}%
Pandya, A.%
\BCBT {}\ \BBA {} Lodha, P.%
\end{APACrefauthors}%
\unskip\
\newblock
\APACrefYearMonthDay{2021}{}{}.
\newblock
{\BBOQ}\APACrefatitle {Social connectedness, excessive screen time during COVID-19 and mental health: a review of current evidence} {Social connectedness, excessive screen time during covid-19 and mental health: a review of current evidence}.{\BBCQ}
\newblock
\APACjournalVolNumPages{Frontiers in Human Dynamics}{3}{}{45}.
\PrintBackRefs{\CurrentBib}

\bibitem [\protect \citeauthoryear {%
Paquet%
, Coulombier%
, Kaiser%
\BCBL {}\ \BBA {} Ciotti%
}{%
Paquet%
\ \protect \BOthers {.}}{%
{\protect \APACyear {2006}}%
}]{%
paquet2006epidemic}
\APACinsertmetastar {%
paquet2006epidemic}%
\begin{APACrefauthors}%
Paquet, C.%
, Coulombier, D.%
, Kaiser, R.%
\BCBL {}\ \BBA {} Ciotti, M.%
\end{APACrefauthors}%
\unskip\
\newblock
\APACrefYearMonthDay{2006}{}{}.
\newblock
{\BBOQ}\APACrefatitle {Epidemic intelligence: a new framework for strengthening disease surveillance in Europe} {Epidemic intelligence: a new framework for strengthening disease surveillance in europe}.{\BBCQ}
\newblock
\APACjournalVolNumPages{Eurosurveillance}{11}{12}{5--6}.
\PrintBackRefs{\CurrentBib}

\bibitem [\protect \citeauthoryear {%
Paul%
, Dredze%
\BCBL {}\ \BBA {} Broniatowski%
}{%
Paul%
\ \protect \BOthers {.}}{%
{\protect \APACyear {2014}}%
}]{%
paul2014twitter}
\APACinsertmetastar {%
paul2014twitter}%
\begin{APACrefauthors}%
Paul, M\BPBI J.%
, Dredze, M.%
\BCBL {}\ \BBA {} Broniatowski, D.%
\end{APACrefauthors}%
\unskip\
\newblock
\APACrefYearMonthDay{2014}{}{}.
\newblock
{\BBOQ}\APACrefatitle {Twitter improves influenza forecasting} {Twitter improves influenza forecasting}.{\BBCQ}
\newblock
\APACjournalVolNumPages{PLoS currents}{6}{}{}.
\PrintBackRefs{\CurrentBib}

\bibitem [\protect \citeauthoryear {%
Philippon%
, Wu%
, Cowling%
\BCBL {}\ \BBA {} Lau%
}{%
Philippon%
\ \protect \BOthers {.}}{%
{\protect \APACyear {2020}}%
}]{%
philippon2020avian}
\APACinsertmetastar {%
philippon2020avian}%
\begin{APACrefauthors}%
Philippon, D\BPBI A.%
, Wu, P.%
, Cowling, B\BPBI J.%
\BCBL {}\ \BBA {} Lau, E\BPBI H.%
\end{APACrefauthors}%
\unskip\
\newblock
\APACrefYearMonthDay{2020}{}{}.
\newblock
{\BBOQ}\APACrefatitle {Avian influenza human infections at the human-animal interface} {Avian influenza human infections at the human-animal interface}.{\BBCQ}
\newblock
\APACjournalVolNumPages{The Journal of Infectious Diseases}{222}{4}{528--537}.
\PrintBackRefs{\CurrentBib}

\bibitem [\protect \citeauthoryear {%
Rech%
}{%
Rech%
}{%
{\protect \APACyear {2007}}%
}]{%
rech2007discovering}
\APACinsertmetastar {%
rech2007discovering}%
\begin{APACrefauthors}%
Rech, J.%
\end{APACrefauthors}%
\unskip\
\newblock
\APACrefYearMonthDay{2007}{}{}.
\newblock
{\BBOQ}\APACrefatitle {Discovering trends in software engineering with google trend} {Discovering trends in software engineering with google trend}.{\BBCQ}
\newblock
\APACjournalVolNumPages{ACM SIGSOFT software engineering notes}{32}{2}{1--2}.
\PrintBackRefs{\CurrentBib}

\bibitem [\protect \citeauthoryear {%
Remongin%
}{%
Remongin%
}{%
{\protect \APACyear {2024}}%
}]{%
remongin}
\APACinsertmetastar {%
remongin}%
\begin{APACrefauthors}%
Remongin, X.%
\end{APACrefauthors}%
\unskip\
\newblock
\APACrefYearMonthDay{2024}{}{}.
\newblock
\APACrefbtitle {{A}vian influenza and export --- agriculture.gouv.fr.} {{A}vian influenza and export --- agriculture.gouv.fr.}
\newblock
\APAChowpublished {\url{https://agriculture.gouv.fr/avian-influenza-and-export}}.
\newblock
\APACrefnote{[Accessed 10-01-2024]}
\PrintBackRefs{\CurrentBib}

\bibitem [\protect \citeauthoryear {%
Robertson%
\ \BBA {} Yee%
}{%
Robertson%
\ \BBA {} Yee%
}{%
{\protect \APACyear {2016}}%
}]{%
robertson2016avian}
\APACinsertmetastar {%
robertson2016avian}%
\begin{APACrefauthors}%
Robertson, C.%
\BCBT {}\ \BBA {} Yee, L.%
\end{APACrefauthors}%
\unskip\
\newblock
\APACrefYearMonthDay{2016}{}{}.
\newblock
{\BBOQ}\APACrefatitle {Avian influenza risk surveillance in North America with online media} {Avian influenza risk surveillance in north america with online media}.{\BBCQ}
\newblock
\APACjournalVolNumPages{PloS one}{11}{11}{e0165688}.
\PrintBackRefs{\CurrentBib}

\bibitem [\protect \citeauthoryear {%
Sanh%
}{%
Sanh%
}{%
{\protect \APACyear {2019}}%
}]{%
sanh2019}
\APACinsertmetastar {%
sanh2019}%
\begin{APACrefauthors}%
Sanh, V.%
\end{APACrefauthors}%
\unskip\
\newblock
\APACrefYearMonthDay{2019}{}{}.
\newblock
{\BBOQ}\APACrefatitle {DistilBERT, a distilled version of BERT: smaller, faster, cheaper and lighter} {Distilbert, a distilled version of bert: smaller, faster, cheaper and lighter}.{\BBCQ}
\newblock
\APACjournalVolNumPages{arXiv preprint arXiv:1910.01108}{}{}{}.
\PrintBackRefs{\CurrentBib}

\bibitem [\protect \citeauthoryear {%
Santillana%
\ \protect \BOthers {.}}{%
Santillana%
\ \protect \BOthers {.}}{%
{\protect \APACyear {2015}}%
}]{%
santillana2015combining}
\APACinsertmetastar {%
santillana2015combining}%
\begin{APACrefauthors}%
Santillana, M.%
, Nguyen, A\BPBI T.%
, Dredze, M.%
, Paul, M\BPBI J.%
, Nsoesie, E\BPBI O.%
\BCBL {}\ \BBA {} Brownstein, J\BPBI S.%
\end{APACrefauthors}%
\unskip\
\newblock
\APACrefYearMonthDay{2015}{}{}.
\newblock
{\BBOQ}\APACrefatitle {Combining search, social media, and traditional data sources to improve influenza surveillance} {Combining search, social media, and traditional data sources to improve influenza surveillance}.{\BBCQ}
\newblock
\APACjournalVolNumPages{PLoS computational biology}{11}{10}{e1004513}.
\PrintBackRefs{\CurrentBib}

\bibitem [\protect \citeauthoryear {%
Seeger%
, Hagerman%
, Johnson%
, Pendell%
\BCBL {}\ \BBA {} Marsh%
}{%
Seeger%
\ \protect \BOthers {.}}{%
{\protect \APACyear {2021}}%
}]{%
seeger2021poultry}
\APACinsertmetastar {%
seeger2021poultry}%
\begin{APACrefauthors}%
Seeger, R\BPBI M.%
, Hagerman, A\BPBI D.%
, Johnson, K\BPBI K.%
, Pendell, D\BPBI L.%
\BCBL {}\ \BBA {} Marsh, T\BPBI L.%
\end{APACrefauthors}%
\unskip\
\newblock
\APACrefYearMonthDay{2021}{}{}.
\newblock
{\BBOQ}\APACrefatitle {When poultry take a sick leave: Response costs for the 2014--2015 highly pathogenic avian influenza epidemic in the USA} {When poultry take a sick leave: Response costs for the 2014--2015 highly pathogenic avian influenza epidemic in the usa}.{\BBCQ}
\newblock
\APACjournalVolNumPages{Food Policy}{102}{}{102068}.
\PrintBackRefs{\CurrentBib}

\bibitem [\protect \citeauthoryear {%
Su%
, Venkat%
, Yadav%
, Puglisi%
\BCBL {}\ \BBA {} Fodeh%
}{%
Su%
\ \protect \BOthers {.}}{%
{\protect \APACyear {2021}}%
}]{%
su2021}
\APACinsertmetastar {%
su2021}%
\begin{APACrefauthors}%
Su, Y.%
, Venkat, A.%
, Yadav, Y.%
, Puglisi, L\BPBI B.%
\BCBL {}\ \BBA {} Fodeh, S\BPBI J.%
\end{APACrefauthors}%
\unskip\
\newblock
\APACrefYearMonthDay{2021}{May}{}.
\newblock
{\BBOQ}\APACrefatitle {Twitter-based analysis reveals differential COVID-19 concerns across areas with socioeconomic disparities} {Twitter-based analysis reveals differential covid-19 concerns across areas with socioeconomic disparities}.{\BBCQ}
\newblock
\APACjournalVolNumPages{Comput Biol Med}{132}{}{104336}.
\newblock
\begin{APACrefDOI} \doi{10.1016/J.COMPBIOMED.2021.104336} \end{APACrefDOI}
\PrintBackRefs{\CurrentBib}

\bibitem [\protect \citeauthoryear {%
USDA%
}{%
USDA%
}{%
{\protect \APACyear {{\protect \bibnodate {}}}}%
}]{%
usdaAvianInfluenza}
\APACinsertmetastar {%
usdaAvianInfluenza}%
\begin{APACrefauthors}%
USDA.%
\end{APACrefauthors}%
\unskip\
\newblock
\APACrefYearMonthDay{{\protect \bibnodate {}}}{}{}.
\newblock
\APACrefbtitle {{A}vian influenza outbreaks reduced egg production, driving prices to record highs in 2022 --- ers.usda.gov.} {{A}vian influenza outbreaks reduced egg production, driving prices to record highs in 2022 --- ers.usda.gov.}
\newblock
\APAChowpublished {\url{https://www.ers.usda.gov/data-products/chart-gallery/gallery/chart-detail/?chartId=105576}}.
\newblock
\APACrefnote{[Accessed 10-01-2024]}
\PrintBackRefs{\CurrentBib}

\bibitem [\protect \citeauthoryear {%
{USDA}%
}{%
{USDA}%
}{%
{\protect \APACyear {2023}}%
}]{%
USDA}
\APACinsertmetastar {%
USDA}%
\begin{APACrefauthors}%
{USDA}.%
\end{APACrefauthors}%
\unskip\
\newblock
\APACrefYearMonthDay{2023}{}{}.
\newblock
\APACrefbtitle {USDA Animal Disease Information: Avian Influenza.} {Usda animal disease information: Avian influenza.}
\newblock
\begin{APACrefURL} \url{https://www.aphis.usda.gov/aphis/ourfocus/animalhealth/animal-disease-information/avian/avian-influenza/2022-hpai} \end{APACrefURL}
\newblock
\APACrefnote{Accessed: Date}
\PrintBackRefs{\CurrentBib}

\bibitem [\protect \citeauthoryear {%
{WAHIS}%
}{%
{WAHIS}%
}{%
{\protect \APACyear {2023}}%
}]{%
WAHIS}
\APACinsertmetastar {%
WAHIS}%
\begin{APACrefauthors}%
{WAHIS}.%
\end{APACrefauthors}%
\unskip\
\newblock
\APACrefYearMonthDay{2023}{}{}.
\newblock
\APACrefbtitle {World Animal Health Information System.} {World animal health information system.}
\newblock
\APAChowpublished {\url{http://https://www.woah.org/en/what-we-do/animal-health-and-welfare/disease-data-collection/world-animal-health-information-system//}}.
\PrintBackRefs{\CurrentBib}

\bibitem [\protect \citeauthoryear {%
Xie%
\ \protect \BOthers {.}}{%
Xie%
\ \protect \BOthers {.}}{%
{\protect \APACyear {2023}}%
}]{%
xie2023episodic}
\APACinsertmetastar {%
xie2023episodic}%
\begin{APACrefauthors}%
Xie, R.%
, Edwards, K\BPBI M.%
, Wille, M.%
, Wei, X.%
, Wong, S\BHBI S.%
, Zanin, M.%
\BDBL {}others%
\end{APACrefauthors}%
\unskip\
\newblock
\APACrefYearMonthDay{2023}{}{}.
\newblock
{\BBOQ}\APACrefatitle {The episodic resurgence of highly pathogenic avian influenza H5 virus} {The episodic resurgence of highly pathogenic avian influenza h5 virus}.{\BBCQ}
\newblock
\APACjournalVolNumPages{Nature}{}{}{1--8}.
\PrintBackRefs{\CurrentBib}

\bibitem [\protect \citeauthoryear {%
Yang%
\ \protect \BOthers {.}}{%
Yang%
\ \protect \BOthers {.}}{%
{\protect \APACyear {2022}}%
}]{%
yang2022human}
\APACinsertmetastar {%
yang2022human}%
\begin{APACrefauthors}%
Yang, R.%
, Sun, H.%
, Gao, F.%
, Luo, K.%
, Huang, Z.%
, Tong, Q.%
\BDBL {}others%
\end{APACrefauthors}%
\unskip\
\newblock
\APACrefYearMonthDay{2022}{}{}.
\newblock
{\BBOQ}\APACrefatitle {Human infection of avian influenza A H3N8 virus and the viral origins: A descriptive study} {Human infection of avian influenza a h3n8 virus and the viral origins: A descriptive study}.{\BBCQ}
\newblock
\APACjournalVolNumPages{The Lancet Microbe}{3}{11}{e824--e834}.
\PrintBackRefs{\CurrentBib}

\bibitem [\protect \citeauthoryear {%
Yousefinaghani%
, Dara%
, Mubareka%
\BCBL {}\ \BBA {} Sharif%
}{%
Yousefinaghani%
\ \protect \BOthers {.}}{%
{\protect \APACyear {2021}}%
{\protect \APACexlab {{\protect \BCnt {1}}}}}]{%
yousefinaghani2021prediction}
\APACinsertmetastar {%
yousefinaghani2021prediction}%
\begin{APACrefauthors}%
Yousefinaghani, S.%
, Dara, R.%
, Mubareka, S.%
\BCBL {}\ \BBA {} Sharif, S.%
\end{APACrefauthors}%
\unskip\
\newblock
\APACrefYearMonthDay{2021{\protect \BCnt {1}}}{}{}.
\newblock
{\BBOQ}\APACrefatitle {Prediction of COVID-19 waves using social media and Google search: a case study of the US and Canada} {Prediction of covid-19 waves using social media and google search: a case study of the us and canada}.{\BBCQ}
\newblock
\APACjournalVolNumPages{Frontiers in public health}{9}{}{656635}.
\PrintBackRefs{\CurrentBib}

\bibitem [\protect \citeauthoryear {%
Yousefinaghani%
, Dara%
, Mubareka%
\BCBL {}\ \BBA {} Sharif%
}{%
Yousefinaghani%
\ \protect \BOthers {.}}{%
{\protect \APACyear {2021}}%
{\protect \APACexlab {{\protect \BCnt {2}}}}}]{%
covidprediction}
\APACinsertmetastar {%
covidprediction}%
\begin{APACrefauthors}%
Yousefinaghani, S.%
, Dara, R.%
, Mubareka, S.%
\BCBL {}\ \BBA {} Sharif, S.%
\end{APACrefauthors}%
\unskip\
\newblock
\APACrefYearMonthDay{2021{\protect \BCnt {2}}}{}{}.
\newblock
{\BBOQ}\APACrefatitle {Prediction of COVID-19 waves using social media and Google search: a case study of the US and Canada} {Prediction of covid-19 waves using social media and google search: a case study of the us and canada}.{\BBCQ}
\newblock
\APACjournalVolNumPages{Frontiers in public health}{9}{}{656635}.
\PrintBackRefs{\CurrentBib}

\bibitem [\protect \citeauthoryear {%
Yousefinaghani%
, Dara%
, Poljak%
, Bernardo%
\BCBL {}\ \BBA {} Sharif%
}{%
Yousefinaghani%
\ \protect \BOthers {.}}{%
{\protect \APACyear {2019}}%
}]{%
yousefinaghani2019assessment}
\APACinsertmetastar {%
yousefinaghani2019assessment}%
\begin{APACrefauthors}%
Yousefinaghani, S.%
, Dara, R.%
, Poljak, Z.%
, Bernardo, T\BPBI M.%
\BCBL {}\ \BBA {} Sharif, S.%
\end{APACrefauthors}%
\unskip\
\newblock
\APACrefYearMonthDay{2019}{}{}.
\newblock
{\BBOQ}\APACrefatitle {The assessment of Twitter’s potential for outbreak detection: avian influenza case study} {The assessment of twitter’s potential for outbreak detection: avian influenza case study}.{\BBCQ}
\newblock
\APACjournalVolNumPages{Scientific reports}{9}{1}{18147}.
\PrintBackRefs{\CurrentBib}

\bibitem [\protect \citeauthoryear {%
Zhang%
\ \protect \BOthers {.}}{%
Zhang%
\ \protect \BOthers {.}}{%
{\protect \APACyear {2022}}%
}]{%
zhang2022intelligent}
\APACinsertmetastar {%
zhang2022intelligent}%
\begin{APACrefauthors}%
Zhang, Y.%
, Chen, K.%
, Weng, Y.%
, Chen, Z.%
, Zhang, J.%
\BCBL {}\ \BBA {} Hubbard, R.%
\end{APACrefauthors}%
\unskip\
\newblock
\APACrefYearMonthDay{2022}{}{}.
\newblock
{\BBOQ}\APACrefatitle {An intelligent early warning system of analyzing Twitter data using machine learning on COVID-19 surveillance in the US} {An intelligent early warning system of analyzing twitter data using machine learning on covid-19 surveillance in the us}.{\BBCQ}
\newblock
\APACjournalVolNumPages{Expert systems with applications}{198}{}{116882}.
\PrintBackRefs{\CurrentBib}

\bibitem [\protect \citeauthoryear {%
Zhao%
\ \protect \BOthers {.}}{%
Zhao%
\ \protect \BOthers {.}}{%
{\protect \APACyear {2019}}%
}]{%
zhao2019semiaquatic}
\APACinsertmetastar {%
zhao2019semiaquatic}%
\begin{APACrefauthors}%
Zhao, P.%
, Sun, L.%
, Xiong, J.%
, Wang, C.%
, Chen, L.%
, Yang, P.%
\BDBL {}others%
\end{APACrefauthors}%
\unskip\
\newblock
\APACrefYearMonthDay{2019}{}{}.
\newblock
{\BBOQ}\APACrefatitle {Semiaquatic mammals might be intermediate hosts to spread avian influenza viruses from avian to human} {Semiaquatic mammals might be intermediate hosts to spread avian influenza viruses from avian to human}.{\BBCQ}
\newblock
\APACjournalVolNumPages{Scientific Reports}{9}{1}{11641}.
\PrintBackRefs{\CurrentBib}

\end{thebibliography}
\end{document}